\documentclass[10pt,journal,compsoc]{IEEEtran}

\usepackage{graphicx}
\usepackage[inline]{enumitem}
\usepackage[dvipsnames]{xcolor}
\usepackage{pgfplots}
\pgfplotsset{compat=1.15}
\usepackage{url}
\usepackage{framed}
\usepackage{arydshln}
\usepackage{flushend}
\usepackage{lipsum} 
\usepackage[numbers]{natbib}
\usepackage{booktabs}

\newcommand{\highlight}[1]{\vspace{2mm} \noindent \textit{#1}}

\newcommand{\concept}[1]{\textsc{#1}}

\newcommand{\hist}[1]{(\raisebox{-1mm}{\includegraphics[height=4.5mm]{img/survey/#1.pdf}})}
\newcommand{\histt}[1]{\raisebox{-1mm}{\includegraphics[height=4.5mm]{img/survey/#1.pdf}}}

\newcommand{\tsetwo}[1]{\textcolor{black}{#1}}

\begin{document}

\title{How Developers Engineer Test Cases:\\An Observational Study}

\author{\IEEEauthorblockN{Maurício Aniche\IEEEauthorrefmark{1}, Christoph Treude\IEEEauthorrefmark{2}, Andy Zaidman\IEEEauthorrefmark{1}\\}
	\IEEEauthorblockA{\IEEEauthorrefmark{1}Delft University of Technology - The Netherlands\\}
	\IEEEauthorblockA{\IEEEauthorrefmark{2}University of Melbourne - Australia\\}
	\IEEEauthorblockA{m.f.aniche@tudelft.nl, christoph.treude@unimelb.edu.au, a.e.zaidman@tudelft.nl}}

\IEEEtitleabstractindextext{
\begin{abstract}

One of the main challenges that developers face when testing their systems lies in engineering test cases that are good enough to reveal bugs. And while our body of knowledge on software testing and automated test case generation is already quite significant, in practice, developers are still the ones responsible for engineering test cases manually. Therefore, understanding the developers' thought- and decision-making processes while engineering test cases is a fundamental step in making developers better at testing software. In this paper, we observe 13 developers thinking-aloud while testing different real-world open-source methods, and use these observations to explain how developers engineer test cases. We then challenge and augment our main findings by surveying 72 software developers on their testing practices. 
We discuss our results from three different angles. First, we propose a general framework that explains how developers reason about testing. Second, we propose and describe in detail the three different overarching strategies that developers apply when testing. Third, we compare and relate our observations with the existing body of knowledge and propose future studies that would advance our knowledge on the topic. 

\end{abstract}

}

\maketitle

\IEEEdisplaynontitleabstractindextext
\IEEEpeerreviewmaketitle

\sloppy

\section{Introduction}

Software testing is an important and challenging software development activity. No wonder large software companies such as 
Google~\cite{whittaker2012google,se-google}, Microsoft~\cite{page2008we}, and Facebook~\cite{facebook} have long been investing
in software testing and ensuring that their developers master different techniques.
However, while we know that testing accounts for a large part of the
software development process, time-wise and cost-wise~\cite{planning2002economic}, many developers do not see
testing as their favourite task~\cite{rettig1991practical} and, worse, do not feel productive when testing~\cite{meyer2014software}.

The difficulty in testing lies in devising test cases that are good enough to reveal bugs. To mitigate this, the software testing research community has been working on different approaches for a long time. The current advances in the areas of random testing~\cite{pacheco2007randoop}, search-based software testing~\cite{mcminn2011search}, or even more recently, neural bug finding~\cite{habib2019neural}, have already been helping developers in finding bugs they cannot do manually. 
However, these tools are not yet able to fully replace humans due to hard problems (e.g., the oracle problem~\cite{barr2014oracle}) that researchers still need to overcome. In practice, this means that software developers are still majorly responsible for testing their software systems.

Since the advent of agile methodologies~\cite{agile-manifesto} and, more specifically, software development methodologies that value technical aspects such as Extreme Programming~\cite{beck2000extreme}, the popularity of (automated) unit testing among software developers has increased drastically.
In fact, 67\% of the 1,112 developers that completed the 14th Annual State of the Agile Report survey~\cite{agile-survey-2020} affirmed to perform unit testing, the most applied engineering practice among the surveyed ones. The idea of ``developer testing'', as suggested by Meszaros~\cite{meszaros2007xunit} and Tarlinder~\cite{tarlinder2016developer}, where developers are also the ones responsible for testing, indeed became a prominent practice.

The popularity of the practice has led to an explosion of books on the topic. A simple search on Amazon for ``software testing'' returns more than 1,000 books. Well-known practitioners such as Fowler~\cite{fowler-test}, Beck~\cite{beck2000extreme,beck2003test}, Martin~\cite{martin2007professionalism}, Freeman and Pryce~\cite{freeman2009growing}, and Hunt and Thomas~\cite{hunt2003pragmatic}, have written about the advantages of automated unit tests as well as how to pragmatically test software systems.
Academics have also been proposing software testing theories, practices, and techniques, in forms of books. Among them, we mention the books of Myers et al.~\cite{myers2011art}, Pezzè and Young~\cite{pezze2008software}, and Mathur~\cite{mathur2013foundations}. These books largely focus on explaining our current body of knowledge on how to perform domain testing, equivalence partitioning, boundary testing, and structural testing (e.g.,~\cite{ostrand1988category,jeng1994simplified,reid1997empirical,hoffman1999boundary,legeard2002automated,samuel2005boundary,zhu1997software}).

While our body of knowledge on software testing is already quite significant, we again argue that developers are still the ones responsible for putting all these techniques together. Therefore, understanding the developers' thought- and decision-making processes on, e.g., how they reason about what test cases to write, which techniques to apply, what types of questions they face when testing, and how they decide it is time to stop, is a fundamental step in making developers better at testing software. 

In this paper, we observe 13 developers thinking-aloud while testing different real-world open source methods, and use these observations to explain \emph{how developers engineer test cases}. We then challenge and augment our main findings by surveying 72 software developers on their testing practices. 

We discuss our results from three different angles. First, we propose a general framework that explains how developers reason about testing. The framework contains six main concepts (i.e., \concept{documentation}, \concept{source code}, \concept{mental model}, \concept{test case}, \concept{test code}, and \concept{adequacy criteria}) and how they relate to each other. We describe how developers leverage each of these concepts when testing. Second, we observe the behaviour of the developers in a holistic way, and propose and describe in detail the three different overarching strategies that developers apply when testing (i.e., guided by documentation, guided by source code, ad-hoc). Third, we compare and relate our observations with the existing body of knowledge and propose future studies that would advance our knowledge in the topic. Finally, we use all the knowledge we acquired to propose recommendations for practitioners (i.e., how they can use the findings of this paper to improve their testing skills), tool makers (i.e., a list of tools that would improve the way developers do testing), and educators (i.e., testing topics that should be taught at university-level).

This paper makes the following contributions:

\begin{itemize}
    \item A general framework and a set of strategies that explain how developers reason about testing, emerged from an observational study with 13 developers and challenged by 72 surveyed software developers. 
    \item A set of actionable guidelines for practitioners, tool-makers, and educators, as well as a research agenda for the empirical software engineering community.
\end{itemize}

\section{Research Method}

The goal of this study is to \emph{understand the thought-process of developers while engineering test cases}. To that aim, we conduct a qualitative study that uses a think-aloud protocol~\cite{fonteyn1993description}. We ask developers with various backgrounds to engineer (automated) test cases for different programs and to verbally explain their thoughts while performing the task. We use their verbal explanations and the video recording of their screens to understand how developers engineer test cases. We then challenge our observations by means of a survey with developers.

\subsection{Study design}
\label{sec:study-design}

Our study design is composed of five steps: 
\begin{enumerate*}[label=(\roman*)]
    \item participants record themselves performing the task (i.e., they engineer automated test cases for a randomly assigned piece of code),
    \item we watch their screen recording, listen to their verbal explanations, and write down our observations,
    \item we qualitatively analyse the observations and iteratively derive a codebook,
    \item we survey a different set of developers to understand how generalizable our findings are.
\end{enumerate*}

We describe each of the steps in the following sub-sections. We share the observations and codebook in our online appendix~\cite{appendix}.

\subsection{Task design}
\label{sec:task-design}

We ask participants to engineer automated test cases for a piece of code that is randomly assigned. We tell participants to follow the same procedure they are used to when writing tests for the software systems they develop in their daily jobs. We ask participants to explain their thoughts verbally while performing the task. As examples, we give participants a list of topics to keep talking about: 
\begin{enumerate*}[label=(\roman*)]
    \item What are you doing right now?
    \item Where did the idea for this test case come from?
    \item Why would you test this?
    \item Why would you not test this?
    \item What challenges are you facing?
    \item What are you not understanding?
    \item What is the next step?
\end{enumerate*}

We send the instructions to participants via e-mail and give them the possibility to send us the recording back within two weeks. While participants are free to start the experiment at any time during these two weeks, we ask them to perform the entire task without stopping.

Each participant is randomly assigned to one program. The random program was selected out of a dataset of programs that we manually devised for this study. The goal of having different programs is to ensure that our findings are less dependent on the program under test.
When selecting candidate programs, we followed three criteria:

\begin{itemize}

    \item Programs have to be of a \emph{familiar domain} to developers. By familiar, we mean a domain that developers with general background in programming can understand. Programs in domains such as mathematics would require developers to also have a background in math, which goes beyond the scope of this paper. After manual exploration, we opt for methods that perform string manipulations. We argue that any developer is familiar with the concept of strings.
    
    \item Programs have to be of \emph{considerable complexity}. Our reasoning is that simple programs might be too easy to test, while too complex programs would require too many hours of work from participants.
    
    \item Classes under test \emph{should not depend on other classes}. While mocking is commonly used among developers~\cite{spadini2017mock,spadini2019mock}, we argue that developers considering whether to use mocks may divert them from our main goal which is to observe how they reflect about deriving test cases.\footnote{We nevertheless believe that observing how developers reason about testing a more complex structure of classes is interesting work, and we suggest it as future work.}
    
\end{itemize}

We selected the Apache Commons Lang\footnote{https://commons.apache.org/lang; the hash of the commit used: e0b474c0d015f89a52.} project, a well-known open-source library that contains, among many other functionalities, methods that manipulate strings. We considered string manipulation a problem that most developers are familiar with, thus matching our first criteria. In terms of complexity, we selected methods that contained $[20,30]$ lines of code, had a cyclomatic complexity of $[5,8]$, and did not depend on any other classes. The thresholds were chosen after manual exploration, in which we observed that methods within that range of complexity would offer a good balance between being complex enough to challenge the participant and yet not too complex so that participants could perform the task in around one hour. From the candidate set of methods, we manually picked four of them. In the following, we describe the methods, based on their official documentation:

\begin{itemize}

    \item \textbf{initials(String str, char... delimiters):} Extracts the initial characters from each word in the String. All first characters after the defined delimiters are returned as a new string. Their case is not changed. The method contains 24 lines of code, five \texttt{if/else} statements, and one \texttt{for} loop.
    
    \item \textbf{leftPad(String str, int size, String padStr):} Left pad a String with a specified String. The string Pad is padded to a given size. The method contains 26 lines of code, six \texttt{if/else} statements, and one \texttt{for} loop.
    
    \item \textbf{rightPad(String str, int size, String padStr):} Right pad a String with a specified String. The string is padded to a given size. The method contains 26 lines of code, six \texttt{if/else} statements, and one \texttt{for} loop.\footnote{The \texttt{leftPad()} and the \texttt{rightPad()} method share similar implementations.}
    
    \item \textbf{substringsBetween(String str, String open, String close):} Searches a String for substrings delimited by a start and end tag, returning all matching substrings in an array. The method contains 29 lines of code, five \texttt{if/else} statements, and one \texttt{while} loop.
 
\end{itemize}

We apply a few transformations to the original code snippet. More specifically, to avoid any bias, we remove parts of the original Javadoc documentation that contained examples of inputs and outputs that could influence the test case engineering process of the developer. We also translated their documentation to Brazilian Portuguese (the main language of all the recruited participants, as we later discuss in Section~\ref{sec:recruitment}) to avoid any confusion from participants that do not fluently speak English. Note that we do not perform any changes in the implementation themselves; the source code is precisely the same as in the Apache library.

Finally, we provide participants with IntelliJ and Eclipse workspaces containing the method they should test in a single class. We ask participants to import the code in the IDE they are more familiar with. We do not give the entire class from the Apache library (which contains several methods that are unrelated to the study), but a new class containing solely the method to test, to avoid confusion. The final version of the code snippets as well as the workspaces can be found in our online appendix~\cite{appendix}.

In the end, participants submit their video recording and the final test class they produced. The videos are used as the main source of analysis. This study was reviewed and approved by the Delft University of Technology's human research ethics council. Participants provided their consent on the participation and were aware that findings would be published in an anonymized way. Our appendix does not contain the raw videos as they contain private information of the participant (e.g., flashes of their e-mail inboxes or messaging apps, or their faces).

\subsection{Recruitment Strategy}
\label{sec:recruitment}

\tsetwo{The goal of our recruitment strategy was to attract professional software developers with experience in writing test cases. Since it is not realistic to conduct random sampling of all software developers who fulfil these criteria, we employed convenience sampling.} We attract participants via social media. The first author of this paper tweeted about the study\footnote{\url{https://twitter.com/mauricioaniche/status/1273285692710432771}, in Brazilian Portuguese.}. In exchange for their participation, we offer a two-hour meet-up on software testing, where participants will have the chance to meet and talk about software testing with the first author of this paper.\footnote{This meet-up happened after the study, and therefore, did not bias the participants.} Given that all the respondents were Brazilians, we decided to allow participants to speak in Brazilian Portuguese, as we conjectured that speaking in their first language would make participants to better focus on tasks and on thinking-aloud.

As a first step in the recruiting process, we ask participants to give us information about their background as developers. We ask the following questions:

\begin{itemize}
    \item Years as a professional software developer.
    \item Years of experience with testing and JUnit.
    \item How often they write automated tests in their current job [Likert scale, 1-5].
    \item Academic background [BSc, MSc, PhD, none].
\end{itemize}
We set as requirements:
\begin{enumerate*}[label=(\roman*)]
\item professional experience with Java, and
\item experience with JUnit and automated test cases.
\end{enumerate*}
\tsetwo{These criteria were essential to filter out developers without testing experience and without experience in our programming language of choice, thus enabling us to capture strategies employed by professional developers. We recommend future work to focus on other languages and testing frameworks, as well as on the challenges encountered by newcomers.}
We received a total of 94 candidate participants in seven days.
We randomly split the participants into the four tasks and send the instructions, as described in Section~\ref{sec:study-design}.

In the end, 13 participants (14\% of those that subscribed to the experiment) fully completed the experiment. We show their demographics in Table~\ref{tab:demographics-full}. We observe that participants have varied levels of experience in software development. The average number of years of experience as a developer is $9.9 \pm 4.0$ (median=10) years, indicating that most of our participants have significant experience in the field. In terms of experience with JUnit, we see an average of $4.0 \pm 3.2$ (median=3) years. 

\subsection{Analysis of the videos and verbal explanations}

Our analysis method contains the following steps: 
\begin{enumerate*}[label=(\roman*)]
    \item we watch the screen recording and listen to the verbal explanations,
    \item we write observation notes that describe the participants' actions, in such a way that it can be coded and understood by any researcher,
    \item we inductively code the observation notes, and finally
    \item we iteratively refine the codebook.
\end{enumerate*}

As a first step, we watch the screen recording and listen to their verbal explanations. The first author of this paper is responsible for this task, given that he is the only one that speaks the language of the participants. The observation notes contain 
\begin{enumerate*}[label=(\roman*)]
    \item the different actions taken by the participant (e.g., ``look at the documentation'', ``write an automated test case'', ``run the test suite'', ``use the debugger''),
    \item the train of thought/reasoning of participants while performing that task, extracted from their verbal explanations (e.g., ``this line can throw a null pointer exception, so it should be tested''),
    \item overall observations of the researcher while observing the participant (e.g., participants writing tests that should clearly fail due to a misunderstanding of the requirements of the program).
\end{enumerate*}
The researcher aimed to be as systematic as possible in this process. To ensure the quality of the observations, the researcher watched each video at least twice. 
The second author of this paper randomly selected three participants and confirmed the appropriateness of the observations extracted from the videos.

The observation notes document is then used for coding. The coding process was mostly conducted by the first author of this paper as he had all the knowledge accumulated from watching the participants. The codes were derived iteratively. The researcher assigned each piece of information in the observation notes to an open code. The researcher also performed axial coding iteratively by finding connections and relationships between the codes. To ensure reliability, once the entire open and axial coding were done, the researcher revisited the codes of all participants.
The final observations that emerged out of the codebook were then revisited and discussed by the three authors of this paper.

We note that the entire process was done iteratively. The researcher would watch the video of a participant, write down the observations, perform open coding, perform axial coding, and then move to the next participant. We argue that doing the analysis iteratively was a good methodological decision as the codes that have emerged from previous participants increased the knowledge of the researcher regarding the phenomenon under study. 

The final document we used for the qualitative analysis contained 37 pages and around 14,400 words, transcribed after 8 hours and 14 minutes of videos. Moreover, to objectively describe the test suites they produced, the table shows the number of test cases, and the branch and mutation coverage achieved by the test suite.
The final version of the codebook is presented in the results section of this paper.

\begin{table*}
\caption{The demographic information of the participants in the think-aloud study (N=13), their assigned tasks, video length, and branch and mutation coverage achived by their test suites. Participants are grouped by their assigned tasks.}
\label{tab:demographics-full}
\resizebox{\textwidth}{!}{\begin{tabular}{lrrrcllrrr}
\toprule
            & \textbf{Experience as} & \textbf{Testing} & \textbf{How often} & \textbf{Highest}           &                   &         &            &          &          \\
\textbf{} & \textbf{developer}     & \textbf{w/ JUnit}           & \textbf{write tests}     & \textbf{academic}  &  \textbf{Assigned}                 & \textbf{Video}   & \textbf{Number of}  & \textbf{Branch}   & \textbf{Mutation} \\
\textbf{ID}          & \textbf{(in years)}    & \textbf{(in years)}      & \textbf{{[}1-5{]}\textsuperscript{b}}        & \textbf{degree} & \textbf{task}              & \textbf{length}  & \textbf{test cases} & \textbf{coverage} & \textbf{coverage} \\ \midrule
P1           & 9             & 7               & 5                & BSc       & Initials          & 0:56:34 & 11         & 100\%    & 92\%     \\
P3           & 8             & 3               & 5                & PhD       & Initials          & 0:30:15 & 6          & 91\%     & 76\%     \\
P7           & 15            & 4               & 2                & BSc       & Initials          & 0:40:53 & 12         & 100\%    & 84\%     \\
P10          & 15            & 1               & 5                & BSc       & Initials          & 0:26:15 & 13         & 91\%     & 92\%     \\
\hdashline
P4           & 14            & 10              & 4                & BSc       & LeftPad           & 0:32:29 & 7          & 100\%    & 88\%     \\
P6           & 11            & 1               & 4                & BSc       & LeftPad           & 0:37:40 & 10         & 100\%    & 88\%     \\
P8\textsuperscript{a}           & 1             & 1               & 3                & None      & LeftPad           & 0:20:02 & 4          & 50\%     & 35\%     \\
P9           & 12            & 6               & 1                & BSc       & LeftPad           & 1:04:27 & 12         & 100\%    & 88\%     \\
P11          & 1             & 1               & 5                & None      & LeftPad           & 0:57:44 & 10         & 100\%    & 88\%     \\
P12          & 14            & 10              & 5                & MSc       & LeftPad           & 0:12:54 & 8          & 100\%    & 88\%     \\ \hdashline
P5           & 10            & 3               & 5                & None      & RightPad          & 1:07:45 & 11         & 100\%    & 88\%     \\ \hdashline
P2           & 10            & 1.5             & 2                & BSc       & SubstringsBetween & 0:27:54 & 7          & 93\%     & 68\%     \\
P13\textsuperscript{a}          & 9             & 4               & 4                & BSc       & SubstringsBetween & 0:18:38 & 6          & 75\%     & 62\%    \\
\bottomrule
\multicolumn{10}{l}{\textsuperscript{a} The participant delivered a failing test method, which we commented out to run the mutation testing tool.}\\
\multicolumn{10}{l}{\textsuperscript{b} Likert scale, where 1 means ``Rarely'' and 5 means ``All the time''.}
\end{tabular}}
\end{table*}

\subsection{Survey design}

The validation survey is devised based on the observations from the qualitative analysis. The survey is composed of seven main parts, one for each of the major concepts that emerged during the analysis (i.e., \concept{documentation}, \concept{source code}, \concept{mental model}, \concept{test cases}, \concept{test code}, and \concept{adequacy criterion} -- all explained in the Results section of this paper) and one for the overall testing strategy that developers apply. 

Each of these parts is composed of multiple Likert scale questions where the participant tells us how often s/he practices a specific observation. For example, one of the items in the \concept{Documentation} section states: ``I read the documentation only at the beginning of the implementation, and that is enough for me to write tests''.\footnote{This is a free translation of the item, as the survey was written in the first language of the participants. In fact, all quotes from participants we present later in this paper are freely translated to English.} Participants would then choose an option out of \textit{never}, \textit{rarely}, \textit{occasionaly}, \textit{frequently}, \textit{all the time}, and \textit{I do not know how to answer}. We note the chosen Likert scale options are a common choice when evaluating frequency.
In total, we have eight options for the \concept{Documentation} section, ten for the \concept{Source Code}, three for the \concept{Mental Model}, six for \concept{Test Case}, nine for \concept{Test Code}, and four to \concept{Adequacy Criterion}. All items are compulsory, i.e., participants are required to pick an option for all the items. Moreover, the survey system randomised the order in which the observations are presented to the users, as a way to reduce any possible bias. 

The final part of the survey is about the overall strategy followed by participants. There, we opt for a multiple-choice option, where participants have to pick the strategy they tend to more frequently apply (i.e., using documentation, source code, or mental model as the main source). We also provide participants with an optional open question to explain their reasoning. The open questions were coded in a separate qualitative analysis process.
Finally, to better understand the sample of developers that take part in our survey, we also include the same demographic questions as in the observational study. These questions appear at the beginning of the survey.

We applied different marketing strategies for the survey. First, we emailed the participants that subscribed for the first part of the task but never completed it. After three days, we also shared the survey among our connections in our social networks (LinkedIn\footnote{\url{http://bit.ly/3apKAAE}, in Brazilian Portuguese.} and Twitter\footnote{\url{https://twitter.com/mauricioaniche/status/1356585005502377987}, in Brazilian Portuguese.}) and industrial partners. 

In the end, we received a total of 72 answers. The survey participants had an average of $10.5 \pm 6.6$ (median 10) years of experience in software development, and an average of $5.3 \pm 3.9$ (median 4) years of experience with testing. Moreover, on a scale from 1 to 5 on how often they write tests in their current project, the majority (83.3\%) of participants answered 4 and 5.

\section{A framework on how developers engineer test cases}

\begin{figure}
\centering
  \includegraphics[width=\columnwidth]{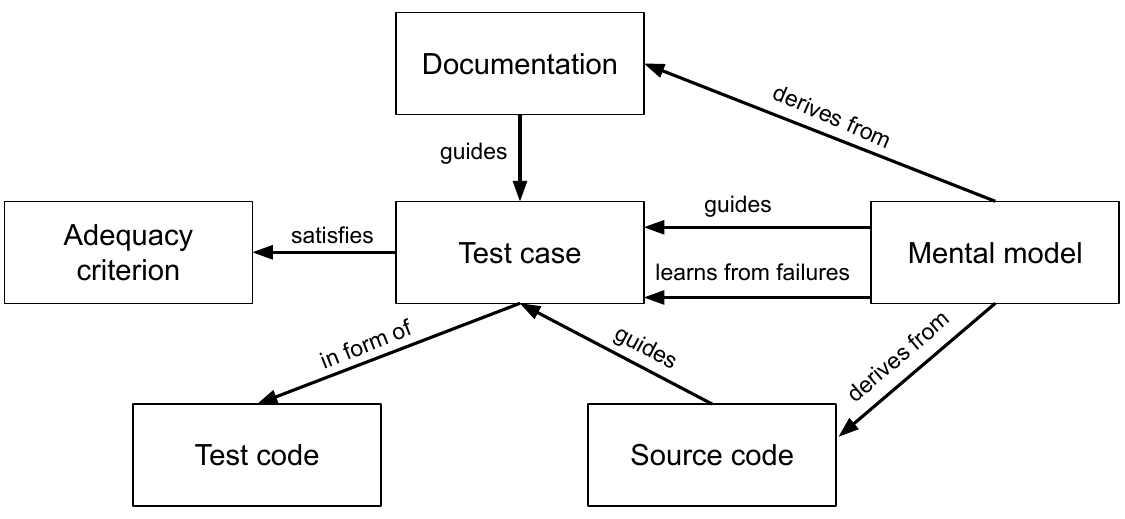}
  \caption{The proposed framework on how developers engineer test cases.}
  \label{fig:framework}
\end{figure}

In this section, we first present the \textit{framework} that emerged out of the results of this paper. Then, we describe the six main concepts that developers utilize when engineering test cases (\concept{test case}, \concept{documentation}, \concept{mental model}, \concept{test code}, \concept{source code}, and \concept{adequacy criterion}).

We introduce the framework in Figure~\ref{fig:framework}:

\begin{enumerate}
    \item A \concept{test case}, the main asset produced by developers during the test phase, describes a set of inputs and a list of expectations. The test case provides the inputs to the program under test and compares the output of the program against the expected behaviour.
    
    \item \concept{Test case}s are derived by a combination of what the developer sees in the \concept{documentation} of the program, the \concept{mental model} of the program that developers build throughout the testing process, as well as what the developer understands from the structure of \concept{source code} of the program under test. 
    
    \item Developers learn from the test failures and use their newly acquired knowledge to improve their \concept{mental model} of the program, which often leads to new \concept{test case}s.
    
    \item Developers automate \concept{test case}s in form of \concept{test code}. 
    
    \item Developers propose new \concept{test case}s until a specific 
    \concept{adequacy criterion} is satisfied. Developers then consider that their testing task is done.
\end{enumerate}

In the following sub-sections, we explain each of these six concepts in detail by means of the data we collected during the observational study and the survey. Whenever presenting evidence from the observational study, we describe the participants from which the observation emerged (e.g., P1 means Participant 1). Whenever presenting evidence from the survey data, we show the percentage of participants that selected one of the items in the Likert scale as well as a miniature bar plot representing the entire distribution of answers. Bar plots contain six bars, in the following order: \textit{I do not know}, \textit{never}, \textit{rarely}, \textit{occasionally}, \textit{frequently}, and \textit{all the time}.

\subsection{Documentation}

The \concept{documentation} of the program under test is used, in one way or another, by all developers (except for P6, who was fully guided by the source code and barely looked at the documentation) as an important source of information. 

We observe developers using the documentation as a way to build an initial \concept{mental model} of the program under test, which is then leveraged as the main source of inspiration for testing during the rest of task.
We use P2 (a very experienced software developer, but with little experience in testing), as an example. As his first action, P2 thoroughly read the documentation of the program (he followed sentence-by-sentence using his mouse pointer, and read it out loud); after finishing reading, P2 then never went back to it, and used solely the \concept{mental model} he had about what the program was supposed to do. P12 (a very experienced software developer and tester) presented a similar behaviour. Interestingly, P12 did not even finish reading the documentation to start with his first test. Once a part of the documentation caught his attention, he jumped straight to the test, and then was mostly guided by his mental model and source code.

Participants also often resorted back to the documentation in moments of confusion and lack of understanding. We observed P1, P5, P7, P8, P10, P11, and P13 revisiting the documentation whenever they did not know what the program was supposed to do in a specific case. P1, for example, after directly experimenting (via test code) with what the program would return in case of input strings with empty spaces and null delimiters, goes back to the documentation to see if there is a mention of the expected behaviour in such cases. P8 experimented with passing a negative number as an input to the program. Before writing the test, he said he would expect the method to either return null or throw an exception. When the program showed a different behaviour, he resorted back to the documentation.

Given that developers tend to resort back to the documentation only when in doubt, one might argue that the quality of the documentation plays a vital role in supporting the developer. While this was not part of the experiment, one participant explicitly mentioned that the quality of the documentation provided in this experiment was high. According to the participant, the documentation provided him with clear business rules that he can use to get started with testing. He says: \textit{``in real life, whenever the ticket\footnote{He used the term ``ticket'' as a reference to tickets in issue trackers, which is often used by software development teams to keep track of the requirements to refer to the documentation.} comes with a nice documentation, testing gets easier''}. 

We also observed two erratic behaviours from developers when using the documentation. P6, in a moment of confusion about what the code had to do, resorted back to the documentation. His main doubt was explicitly mentioned in the documentation; however, he simply could not find it. For some reason, the participant focused his attention on the open text part of the documentation, but did not look at the documentation of the specific input parameter he was having trouble with. In addition, we observed P2, at the end of the task, affirming that he did not test other exceptional cases (in his specific case, passing an empty string to the program) because the specification said that empty strings and null strings are treated the same way; however, he did not ensure the implementation handled it correctly.

\highlight{Survey results.} Documentation also seems to be a popular source of information for testing among the survey participants. Around 37\% of the participants affirm to frequently use the documentation as main source for testing, and 19\% affirm to use it all the time; on the other hand, 16\% of participants use it rarely and only 4\% never use it~\hist{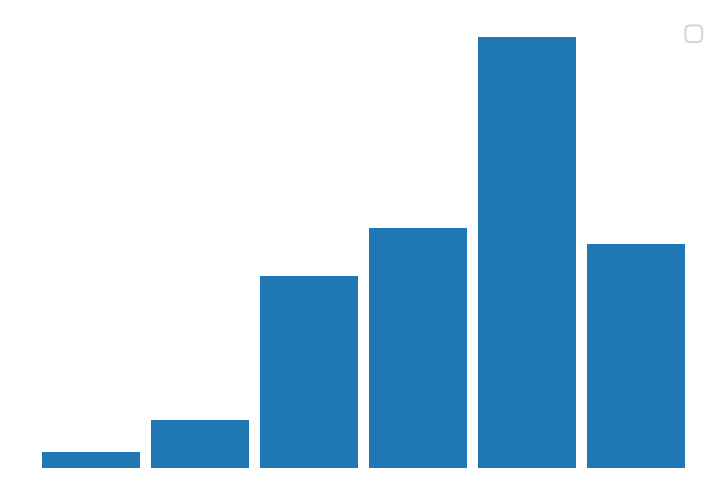}. Participants also disagree that reading it at the beginning only (as done by P2 and P12) is enough to write the tests. 25\% of the participants say this is never possible, and 30\% say this only rarely happens; interestingly, 19\% affirm to do it all the time~\hist{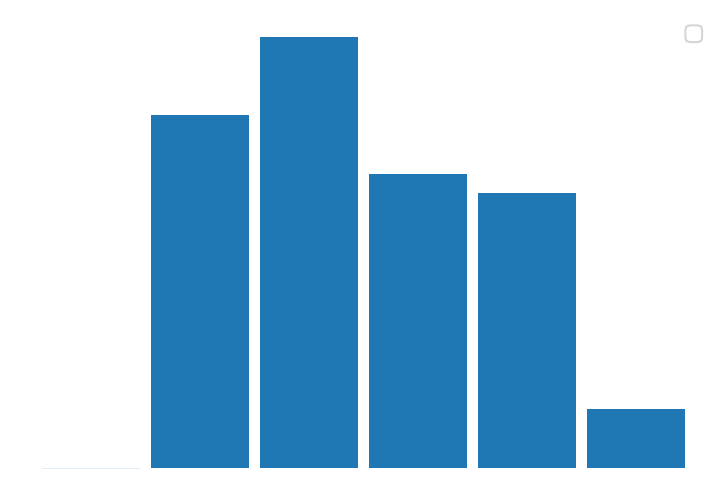}. In fact, revisiting the documentation looking for more tests is a common activity, as 27\%, 38\%, and 20\% of our participants affirm to do it occasionally, frequently, and all the time, respectively~\hist{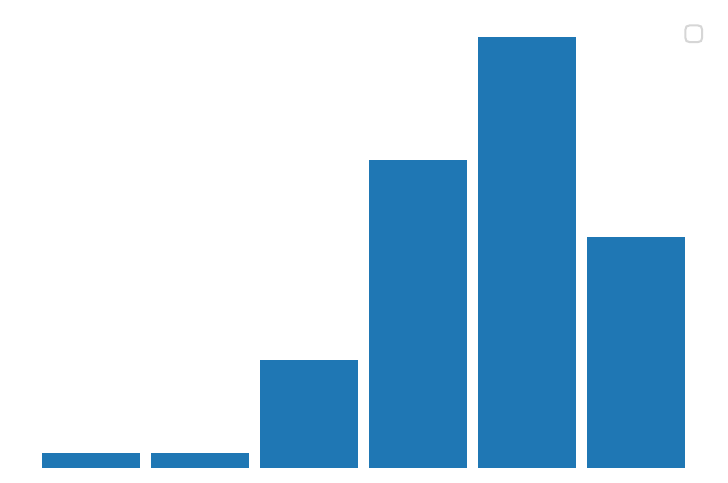}. The quality of the documentation is also something that participants often perceive as not complete in their daily jobs. Participants affirm to stumble across a case where the documentation is not clear about what the program should do occasionally (33\%), frequently (34\%), or all the time (20\%); only 10\% say that this is a rare situation, and only one participant affirmed to never have this problem~\hist{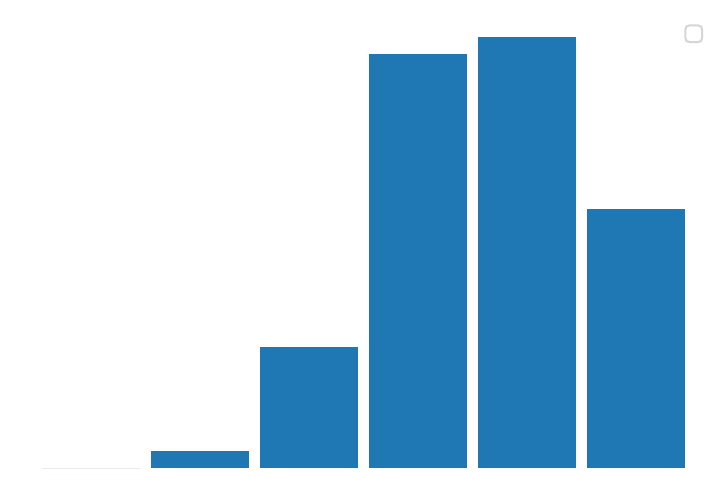}. Finally, the erratic behaviour we observed from P2 does not seem to be common in the wild, as survey participants affirm to not take the current behaviour of the method as correct when the documentation does not talk about the case: 40\% of them affirm they never do it, and 23\% do it only rarely~\hist{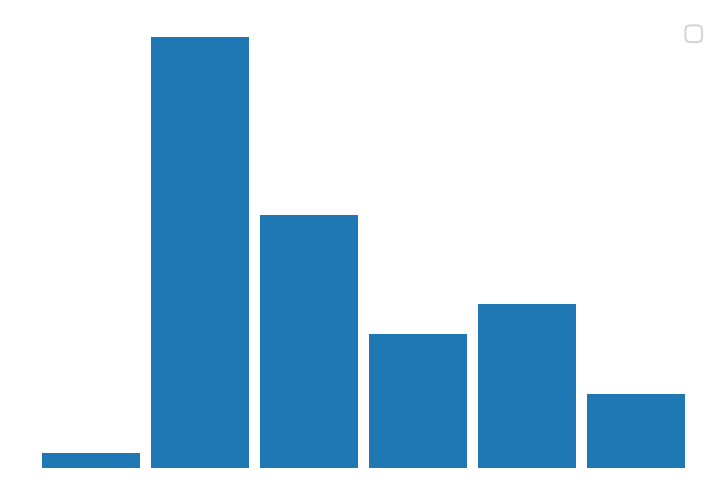}.

\begin{framed}
\noindent Main takeaways (\concept{Documentation}):

\begin{enumerate}[label=(D\arabic*),leftmargin=*]
    \item Developers use the documentation to build an initial mental model of how the program under test works.
    
    \item Developers resort back to the documentation mostly when they lack the knowledge on what to expect from the program in a specific case.
    
    \item Incomplete or unclear documentation might prevent developers from understanding what precisely they have to test for. Moreover, incomplete and unclear documentation seem to be prevalent in the wild. 

\end{enumerate}

\end{framed}

\subsection{Source code}

While source code alone is rarely used as a sole source of information for testing (P6 only), reading and comprehending source code is a prevalent activity when devising tests. Except for P2, P9, and P10, all other participants dived into the source code for comprehension. The number of times that participants resorted back to the source code varies significantly. For example, we observed P11 going back to the source code eight times throughout the task, while P5 went back only once.

The participants' objectives when inspecting the source code highly varied. We observe participants exploring the code in a systematic manner, i.e., line-by-line or branch-by-branch (P3, P11), to understand how the method would behave in a specific case (P1, P4, P6, P7, P11, P12), to get an overall idea of what the implementation is about (P4, P7, P8, P13), to find a way to reach a specific branch (P5), and to understand a test failure and why their mental model/expectations did not match with what the method actually did (P6, P8). 

Interestingly, while code comprehension is a constant activity, not fully comprehending code did not stop some of our participants to continue writing tests (P3, P11). For example, P3 tried to comprehend why the implementation declares an array of characters with a \texttt{strLen/2+1} size, but failed; he just decided to move on and keep building an overall understanding of the method. The same happened with P11; in his case, the participant could not figure out the reason behind the array access \texttt{padLen[i\%strLen]}, and decided to continue comprehending the rest of the code.

Another interesting behaviour we observed was that many participants (P1, P3, P4, P6, P8, P9, P10), after being in doubt about what the method should do for a specific case, accepted the provided output without double-checking in the documentation whether this was indeed the expected behaviour. For example, P1 wrote test methods solely to help him explore the behaviour of the method for a given input. However, he accepted the return of the method as the correct/expected output. P4, after observing the execution result of the method, went to the production code to understand, implementation-wise, why that was the return of method; nevertheless, the participant did not confirm it in the documentation. P6 even vocalized such an action. After the failing of a test, the participant observed JUnit's assertion message (i.e., expected X but returned Y), and said: \textit{``Oh, this is what the method should return.''}. 

The process of comprehending the source code of a program under test is often augmented through the use of a debugger (P2, P4, P5, P6, P11, P13), with some participants making intensive use of the tool (e.g., P6 used the debugger a total of five times). Some participants do not even wait for a test to fail in order to start the debugger (P5, P6, P13). We also observe participants debugging as a way to build an initial understanding of the method at the very beginning of the task (P2), to improve their current understanding of the program (P6, P11), to better understand how to reach a specific branch of the code (P4, P5), or to confirm that the target branch was indeed covered (P6, P11).

As for rare behaviours, P7 was the only participant that performed slight modifications to the production code to better understand it. Somewhat similar, P11 added comments to the production code to improve his understanding of the code.\footnote{We make no conclusions regarding these practices, as participants may be held to believe that they were not supposed to change the production code in this experiment. We discuss that in our threats to validity section.} Finally, we also saw participants resorting to websites like Stack Overflow to better understand some Java constructs or API calls (P3, P7).

\highlight{Survey results.} The survey results confirm our observation that source code is a popular source of information for testing. 25\% of the participants affirm to use it all the time, and 39\% to use it frequently~\hist{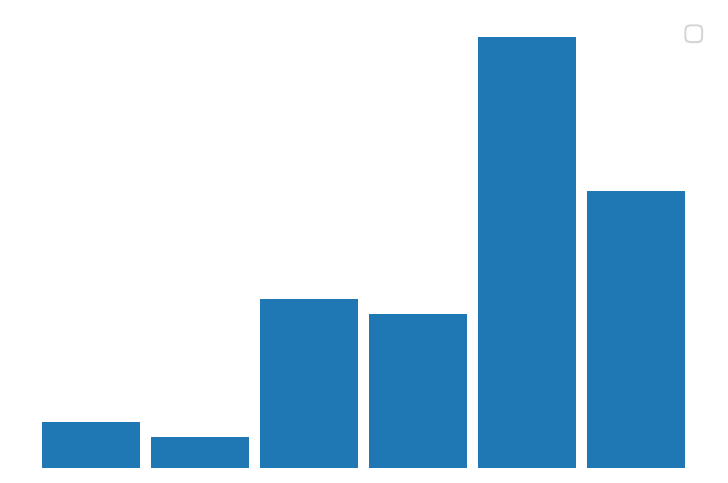}. In fact, if we compare the numbers with the ones presented in the documentation section, we see that source code is slightly more popular. 

The survey also shows that comprehending what the production code does is a very intense activity. The clear majority of participants affirm that a large part of their time is spent on comprehending what the code does~\hist{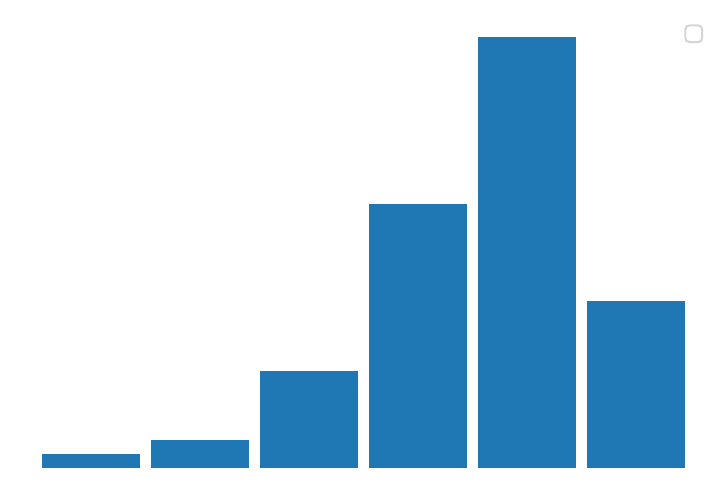}: 16\% affirm that the majority of the time goes to comprehension all the time, and 43\% say frequently. Trying to understand the method under test completely before testing it also seems to be a common behaviour by the developers~\hist{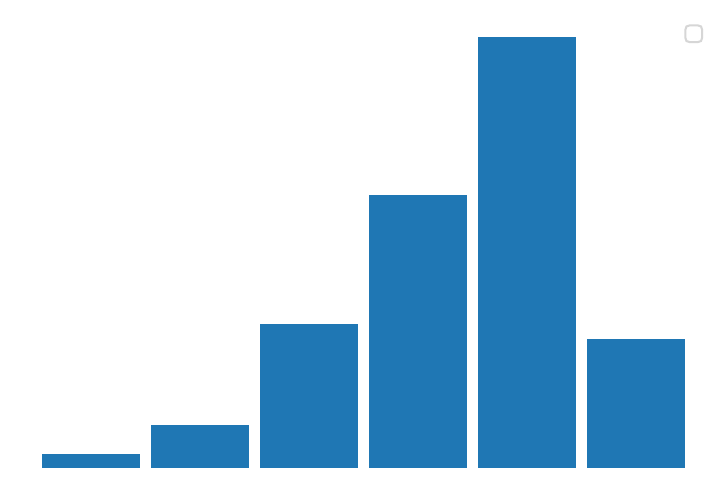}, as 12\% affirm to do it all the time, and 41\% to do it frequently, while 18\% rarely or never do it. On the other hand, testing a method even without full understanding (as we observed in the first part of this study) does not seem to be a default behaviour, as 18\% never did it, and 25\% say they do it only rarely; nevertheless, 25\% still affirm to do it frequently~\hist{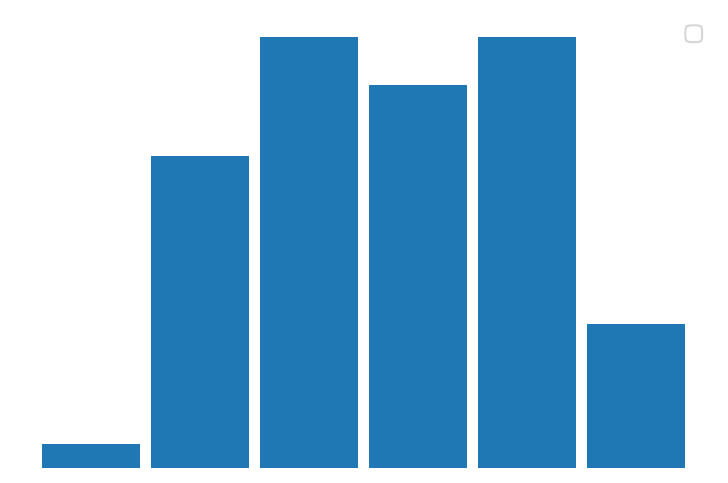}.

Resorting immediately to the documentation once a piece of code is not clear enough~\hist{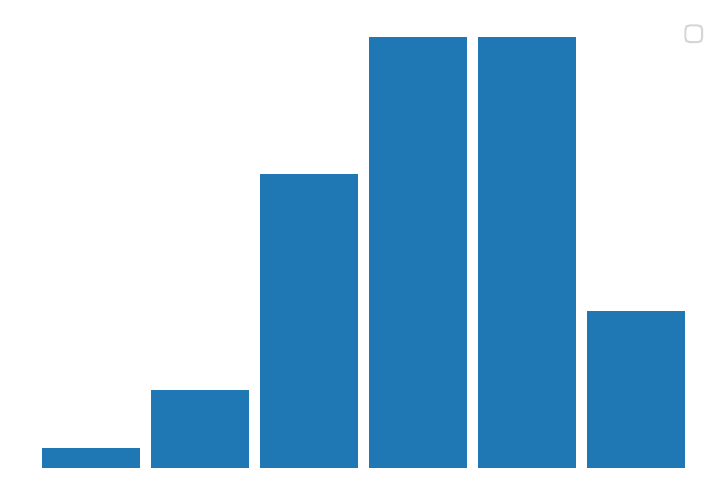} is a popular behaviour, as 11\% and 30\% of the participants either do it all the time or frequently. The use of the debugger to navigate and better understand the code~\hist{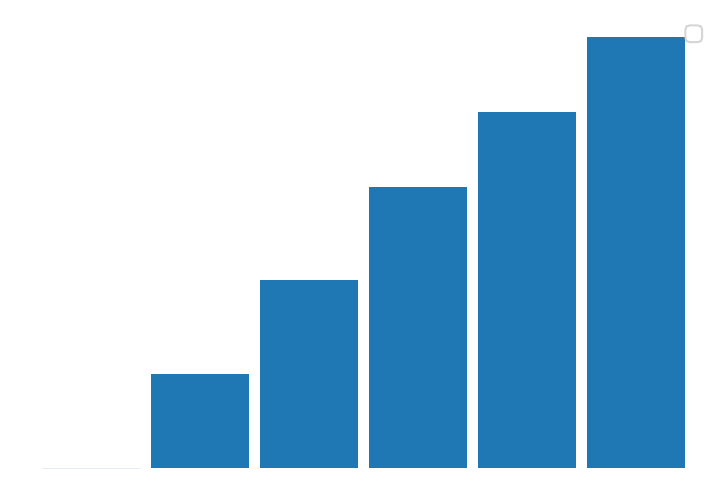} is also highly common, as 32\% and 26\% do it all the time or frequently.
The debugger is also useful to understand how to make the test visit a specific branch of the code~\hist{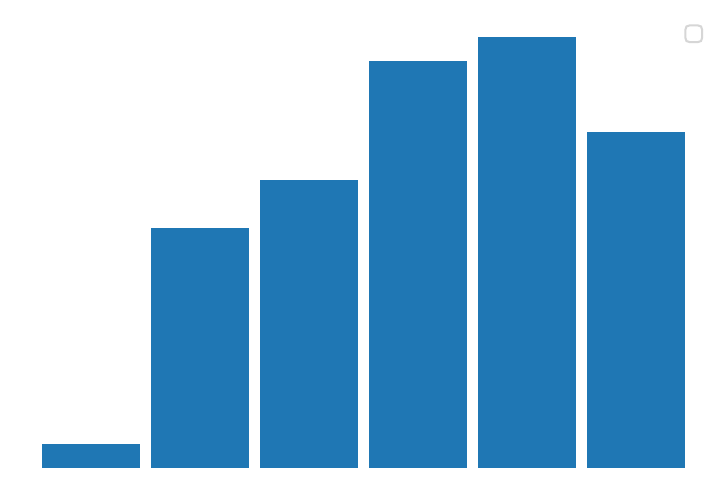}, and 45\% do it all the time or frequently. 
Interestingly, a large number of survey participants affirm to resort to the debugger right away as soon as they have a failing test: 29\% of our participants affirm to do it all the time, and 29\% frequently~\hist{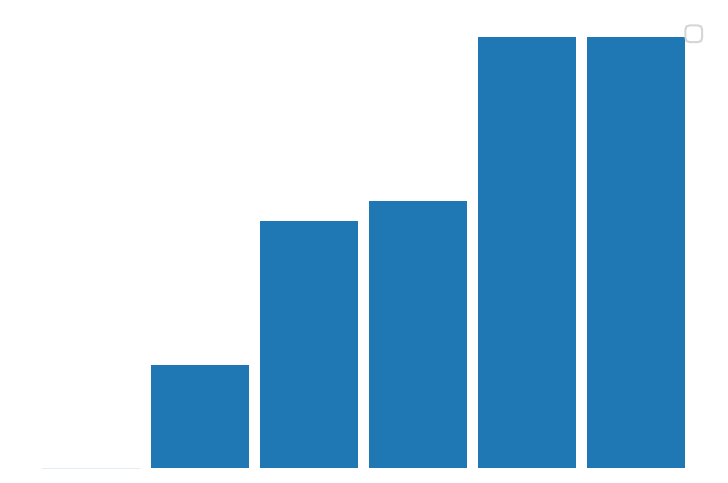}. 
Moreover, the use of code coverage tools as a way to have more ideas on what to test is also popular, with 25\% and 26\% of the survey participants affirming to use them all the time or frequently~\hist{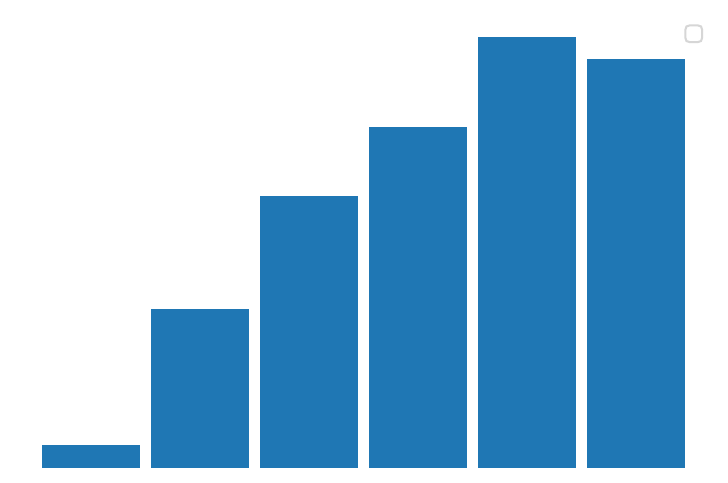}.
Finally, making slight modifications to the production code to better understand it is not a common behaviour, with 37\% of participants never doing it, and 15\% only doing it in rare occasions~\hist{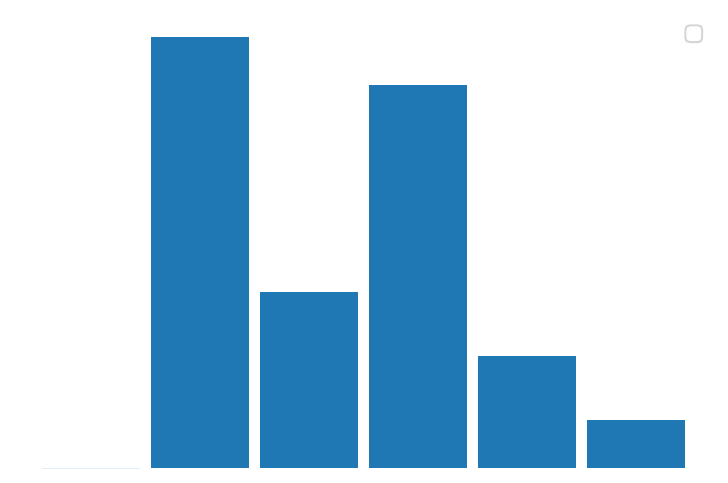}.

\begin{framed}
\noindent Main takeaways (\concept{Source code}):

\begin{enumerate}[label=(S\arabic*),leftmargin=*]

    \item Developers strongly rely on the source code to comprehend the program under test.
    
    \item The (partial) lack of comprehension in parts of the implementation does not prevent developers from continuing their testing activities.
    
    \item Developers often trust the outcomes of a method without double-checking the documentation.
    
    \item The debugger supports developers in better understanding the code as well as to help them understand how to reach specific branches. Resorting to the debugger immediately after a test failure happens is a common behaviour.
    
    \item Developers frequently use code coverage reports to augment their test suites.
        
\end{enumerate}
\end{framed}

\subsection{Mental Model}

The \concept{mental model} that developers build throughout the task plays an important role in the testing process. In the previous sections, we discussed how developers use the documentation and the source code of the program under test to build a mental model of how the program behaves. In this section, we discuss two other important factors that support developers in building and improving their mental models: \emph{hypothesis testing} and \emph{learning from previous failures}.

Throughout the tasks of the 13 participants, we have identified 61 moments where participants were raising hypotheses about the behaviour of the code. Participants would then test those hypotheses (by means of an automated test case), observing the behaviour of the program, and updating their mental model. We provide a few examples to illustrate this idea. P1 knew that whitespaces were considered a special character in the input. He then asked himself: \textit{``is a \texttt{\symbol{92}n} also considered a whitespace [i.e., has the same behaviour as whitespace]?''} Instead of looking at the implementation, the participant simply tested his hypothesis by writing an automated test case and observing its output. P1 later raised another hypothesis: \textit{``what happens if I repeat the delimiter [part of the input]? Actually, what happens if I have a sequenec of delimiters in the input? I have no idea.''} He then tried it out in another test method. P2 questions himself as to whether the program was case sensitive or insensitive. He then wrote a test case based on this hypothesis, and said to himself: \textit{``I believe this will fail.''} The test indeed failed, showing that his mental model was already correct. P5 asked himself: \textit{``what happens if I pass an empty string? Will it return an empty string back?''} P8 hypothesizes: \textit{``If pad string [the input] is null, I expect it to add a space [to the output].''} We observed similar hypotheses in all the participants.
It is interesting to note that developers rarely resort back to the documentation or to the source code to validate their hypotheses. The most common way was to quickly write a unit test that calls the production method with the given input and observe its output. 

Moreover, something we repeatedly observed among participants (especially P1, P3, P5, P6, P7, P8, P12, and P13), was that the first hypotheses and test cases were always focused on ``simple'' test cases, i.e., test cases with basic rather than complex inputs, as to facilitate the process of building an initial mental model. P6, after writing a first test case passing \emph{null} as input, says: \textit{``I made this one simple, so that I get the rhythm.''} P7 says: \textit{``Let's do a first exploratory test. Let me start with a first basic test here.''} P12 says: \textit{``I would start with things I know of how it works.''}

Often, their beliefs were disproven by the test result. In such cases, we observed developers then reflecting on the differences between their conjectures and the concrete output of the model. We observed many ``a-ha moments'' where participants clearly improved their understanding of the program under test. Having a failing test, or a wrong hypothesis is a common behaviour. We have seen all participants learning from failures, with the exception of P3 who never had a failing test. Interestingly, in some cases, participants were able to understand why their conjectures were incorrect right after observing the output of the method. In such cases, participants would simply move to the next test, which was often inspired by the new knowledge they just acquired. In other cases, developers experienced more difficulty in understanding why that was the output. In these cases, developers often resorted back to the documentation and the source code (often via debugging) to precisely understand what was happening. P6, for example, after a quick debug session and documentation and source code reading says: \textit{``Oh, I understood it [the expected behaviour] wrongly. The correct behaviour should be [...]''}. We also observed an interesting behaviour from P7 and P13. After not being able to understand the current failing test, P7 went back to the previous test, and tried to slightly simplify its input in an attempt to isolate the precise behaviour he was not understanding. Similarly, P13 decided to simplify the input in its current test (i.e., from a long string to a short string), hoping that this would help him understand the output. Finally, P2, P4, P10, P11, P12, and P13, faced situations where the test was failing not because they understood it wrongly, but because the test code was wrong; we discuss more about this later.

As for rare behaviours, we note that one participant (P13) was never able to figure out why a test was failing, even after many attempts. At the end, he conjectured that there was a bug in the program under test, which was not the case. We also note that P9, at the beginning of the task, affirmed to be familiar with the program under test. He says: \textit{``I do not know this particular implementation, but we have a similar leftPad() function in our project, so I understand what it does.''}. While we did not notice any clear difference between P9 and the other participants that worked on the left pad task, we conjecture that previous domain knowledge may influence the way participants engineer test cases. We discuss work related to the effects of experience and testing in the related work section of this paper.

\highlight{Survey results.} The survey results show that participants perceive the use of their mental model while testing, although less than other sources of information. 11\% of the participants affirm to use it all the time, and 40\% to use it frequently. 25\% of participants affirm to use it only occasionally, 8\% to rarely use it, and 12\% to never use it~\hist{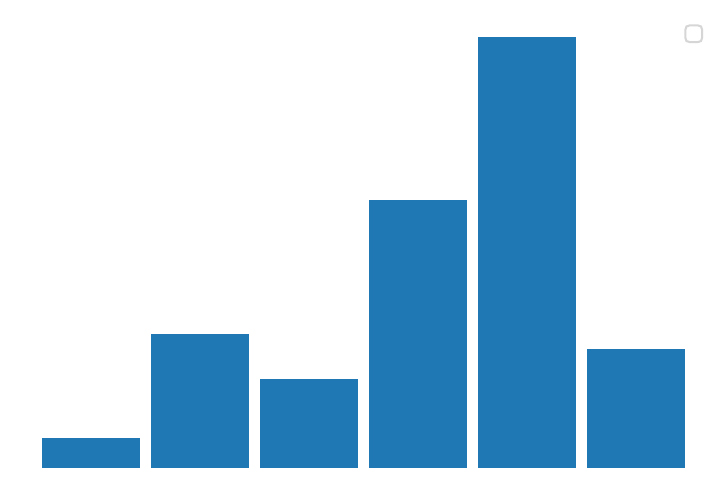}. Interestingly, making use of test code to learn about the program behaviour, instead of reading from the documentation or source code, is less prevalent in the survey than in the observations of the 13 participants. Only 11\% affirm to do it all the time, and 20\% to do it frequently; 31\% affirm to rarely do it, and 11\% to never do it~\hist{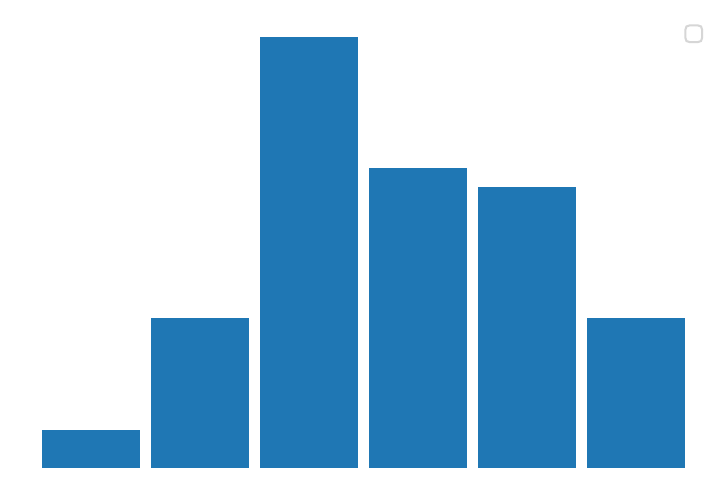}. On the other hand, reflecting about a test failure first, before jumping to the documentation or source code, as noted in the observational study, is a very popular behaviour among our survey participants, with 20\% doing it all the time, and 43\% doing it frequently~\hist{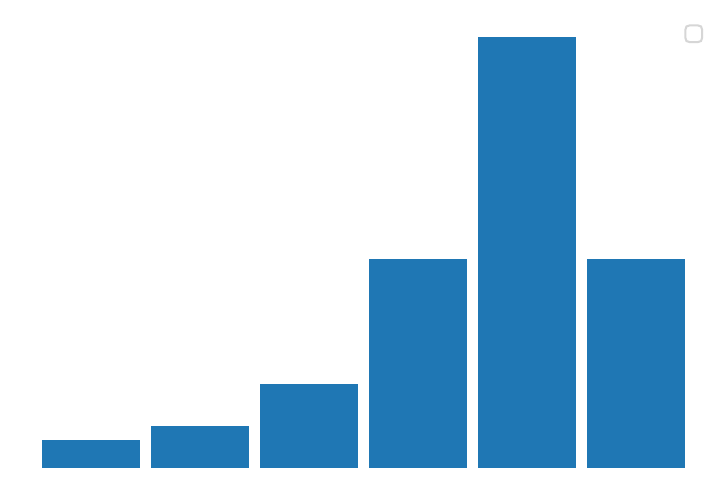}.

\begin{framed}
\noindent Main takeaways (\concept{Mental model}):

\begin{enumerate}[label=(M\arabic*),leftmargin=*]
    \item Although not explicitly noted by developers, the mental model that they build while testing plays a role in the test cases they engineer.
    
    \item The mental model is built not only through the documentation and the source code, but also through hypothesis testing and failures.
    
    \item Developers start building their mental model by exploring simple cases first (i.e., small inputs focused on good weather cases).
\end{enumerate}
\end{framed}

\subsection{Test case}

The \concept{test case}s engineered by the developers are the core of the testing activity. As we saw before, test cases are often derived after what participants observe in the documentation, after the source code, and after the mental model that they build. Across the 13 participants, we observed 32 moments where developers were clearly guided by the structure of the source code (i.e., developers engineered a test after identifying a branch or line that they wanted to exercise), 10 moments clearly guided by the documentation (i.e., developers engineered a test after looking at a specific part of the documentation), and 22 moments where developers were clearly guided by their mental models (i.e., guided by the hypotheses they generated). While quantitatively comparing these numbers is not our main goal, they show that developers make intense use of varied sources of information.

We also observed 48 moments where developers considered testing a boundary or exceptional case. In fact, we observed such tests coming from all participants, but P4 and P13. While this massively varies among the participants (e.g., we counted nine moments for P1, and six moments for P5, P7, and P10, but observed it in only a single moment for P3, in two occasions for P2 and P9, and zero times for P4 and P13), this clearly shows that testing exceptional and boundary cases is part of the developers' testing activities. 

Another interesting characteristic of the engineered test cases is that all of them were composed of simple, short, single, random inputs. For example, string parameters, participants rarely went for strings larger than just a few characters. The same happened with integer inputs, where participants rarely went for large integers.  Interestingly, we observed P1 attempting to create a test case with a very complex string; however, after having some issues in writing the proper assertion for the complex input, he stopped and said: \textit{``That is too complex and it's not really needed [adapted from a slang].''} The participant then went for a simpler input. In addition, we never observed a specific pattern or systematic way of proposing the concrete inputs, i.e., they were all randomly chosen by the developers. Moreover, we notice that developers also tend to pick single inputs for each test case they engineer rather than trying several inputs for the same partition. The only exception being P10 who provides multiple inputs, all equivalent, per test case.

As for the less frequent observations, we observed P9 and P12 purposefully not writing specific test cases. P9 decided not to test extreme integer values (e.g., the maximum value allowed for an integer in Java). He says that, for such a case, he would prefer to wait and see if the problem would really happen in production before writing a test for it. Similarly, P12 argued that he would try out more extreme values if that piece of code belonged to a critical part of the software system, otherwise he would consider it not necessary. 

\highlight{Survey results.} The survey results confirm our observations. First, we see that survey participants indeed aim at testing all the exceptional cases~(\histt{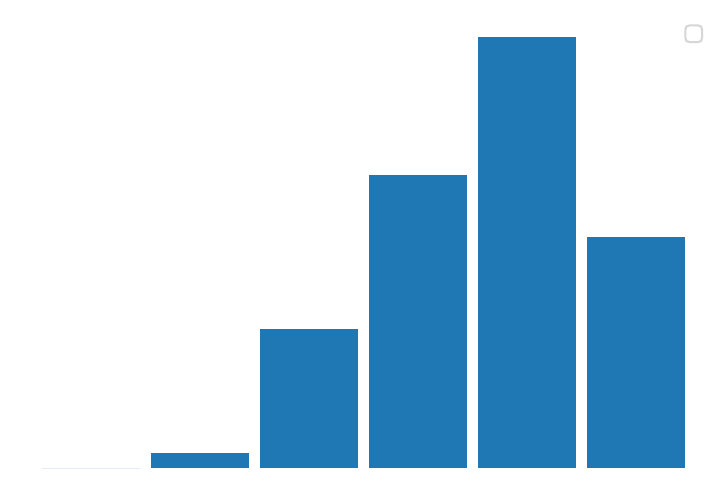}; 20\% all the time, 39\% frequently) and boundary conditions~(\histt{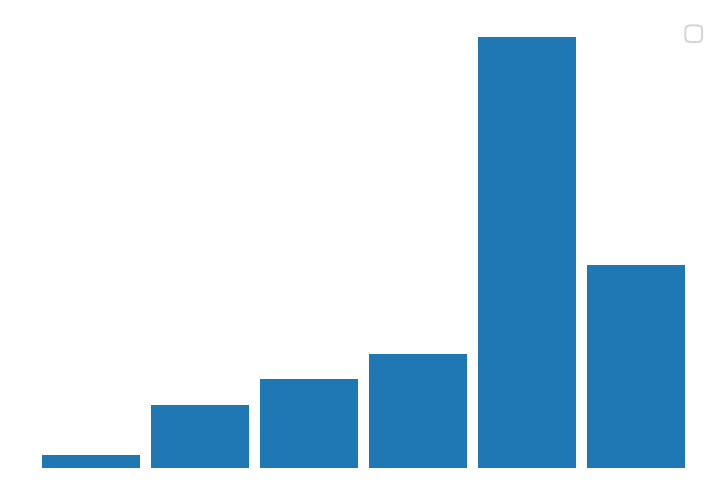}; 22\% all the time, 47\% frequently) they can see. Testing the behaviour of the method with extreme values (e.g., maximum integer value possible) is less popular than others, but still, developers affirm to do it quite often~(\histt{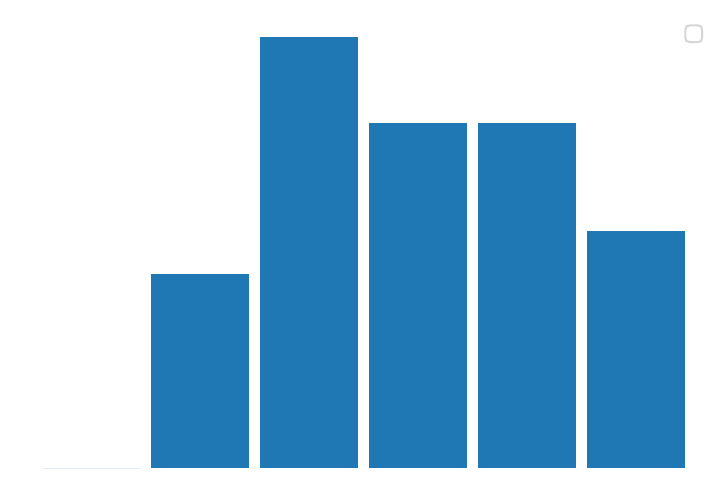}; 15\% all the time, 22\% frequently). Finally, survey participants affirm to opt for short and simple inputs in their test cases, with 11\% doing it all the time, 30\% doing it frequently, and 38\% doing it occasionally~\hist{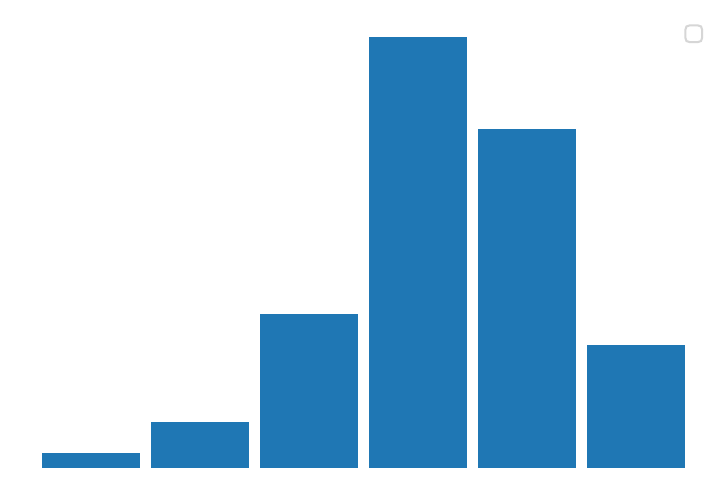}; we nevertheless note that this behaviour was observed in 100\% of the participants in the first part of the study. Finally, while we could not figure out ``from where'' the concrete input values came that participants provided to the test cases, and we assumed they were random, survey participants seem not to fully choose these values randomly. 17\% affirm to never pick them randomly, 26\% to only do it rarely, and 20\% to only do it occasionally~\hist{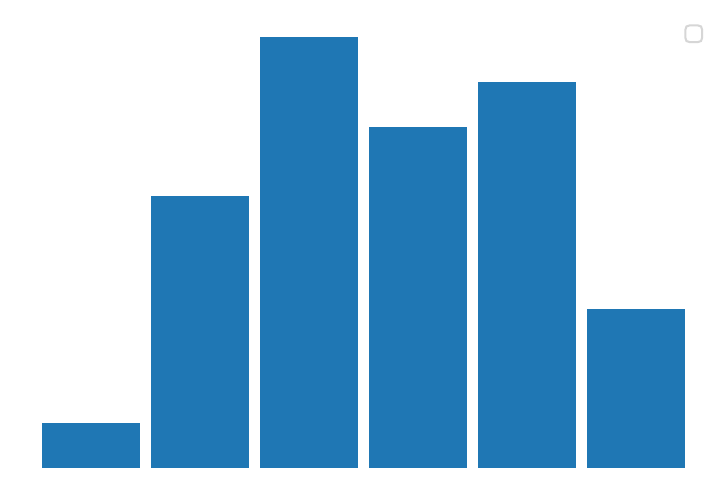}.

\begin{framed}
\noindent Main takeaways (\concept{Test case}):

\begin{enumerate}[label=(TC\arabic*),leftmargin=*]
    \item Test cases are derived from varied sources of information, i.e., documentation, source code, and mental model. 
    
    \item Developers focus not only on ``happy path'' cases, but also on boundary and exceptional cases.
    
    \item The inputs used in test cases tend to be simple, short and random. Test cases contain a single input that exercises the intended behaviour.

\end{enumerate}
\end{framed}

\subsection{Test code}

The test cases that developers engineered (after the documentation, source code, or mental model) were always concretely implemented as \concept{test code} or, in the case of this study, JUnit methods. As such, developers were focusing on producing test code all the time.

The primary observation related to test code is how developers reused previously implemented test methods as a starting point for the next test method. In fact, all participants made use of some sort of code reuse throughout their tasks. After writing the first working test method, participants rarely wrote another test method from scratch; rather, they tended to copy and paste the previous method, modify its name, inputs, and assertions. 

The reusability goes beyond the structure of the test method. Interestingly, participants, more often than not, ``took inspiration'' from the input values defined in the previous test; when devising the next new test case, participants tended to only make slight changes in the previous input, enough for the input to serve for the scenario they wanted to test. For example, P1's first test used the string \texttt{A B C D} as input. His next tests make use of \texttt{Aa Bb Cc Dd} and \texttt{Ab aB AB}; note how the next inputs are highly influenced by the first defined input, i.e., once P1 decided to go for ``A B C D'', other tests were only variations of this initial seed. 

When it comes to the internal code quality of the test code, participants often stopped devising new test cases to refactor the test code. The number of refactorings varied among participants (ranging from zero refactorings from P2, P3 and P8 up to five refactoring moments from P11 and eight refactoring moments from P9). We observed different types of refactorings, i.e., variable renaming (P1, P4, P6, P10), extracting variables (P5, P6, P9; although P5 later decided to rollback the refactoring as he believed that the repetition, while bad, increased legibility of the test code), adding code comments in the test (P5, P11; although P11 later removed the code comments), method renaming (P5, P6, P7, P9, P11, P12, P13), and statically importing libraries as to make method calls shorter (P9, P11). We note the emphasis on renaming the test methods. We observe that, after having a clear understanding of what the test case should be about, participants often refined the name of their test method to better explain what that test case was about. We highlight P9 who engaged in method renaming several times throughout the task. His test methods were initially called \texttt{t1()}, \texttt{t2()} and \texttt{t3()}. At some point (we conjecture once the participant had a better understanding of the problem), he started to look for better names for the tests. Test \texttt{t1()}, for example, first became ``should Return The Same String When Size Is Equals To String Size''\footnote{Translated literally to English. Spaces added just to improve the readability of the text.}, and then later was changed to ``should Return The Same String When Size Is Equals To Number of Characters In Str'', as the participant felt that ``number of characters in the string'' was more specific than ``string size''.

Test code may contain bugs, and we indeed observed tests failing not due to wrong understanding of the program, but due to a bug in the test code itself (P2, P4, P10, P11, P12, P13). P2 wrote a test that threw an exception due to bad coding; the participant mistakenly understood that as if the program did not behave as he expected. After some time, the participant found the real cause of the failure. Forgetting to update the new test code after copying and pasting the previous one happened for three of the participants (P4, P10, P13). P4 and P13 forgot to change the assertion, which let the test fail for wrong reasons. P10 made this mistake three times. The first time, he also forgot to change something in the test that was copied and pasted from the previous one. The second time, the participant managed to make the test pass, without writing the precise assertion for that test. The third time, the participant was never able to identify that the cause of the bug was the test code. P11 made a slight mistake in the assertion (the expected output should have been ``BBAA'', but he wrote ``AABB`` instead), a mistake which he quickly identified. 

As for other behaviours, we observed developers reflecting about whether two tests were duplicated or not (P7, P11), developers deleting previously created tests (P7), making use of a single test method for more than one test case (P10, P12), reorganizing the order that tests appear in the test class as to put methods that exercise similar behaviour closer to each other (P9, P10), weak assertions rather than strong assertions (P2), wrong usage of the testing framework (P10, P11) or searching on the internet for specific features in the framework (P5, P8, P11), making use of a print statement to print the output of the method (P8, P9, P13), and to simplify the input of a previously working test case (P7, P13).

\highlight{Survey results.} Our survey results show that copying and pasting a previous test method as a way to start the next one is indeed a common behaviour~(\histt{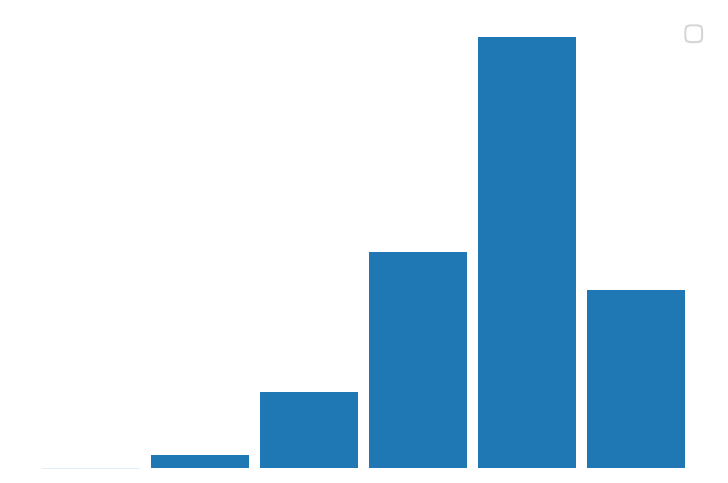}; 19\% all the time, 47\% frequently). On the other hand, participants affirm to not use the same base/seed for the inputs of their tests~(\histt{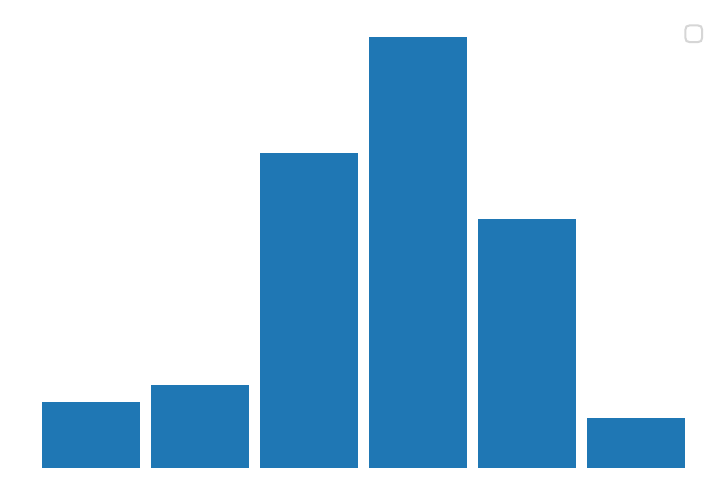}; 36\% occasionally, 26\% rarely, 7\% never). This somewhat contradicts our observation in the first part of this study where this happened more often. Test code refactoring also seems to be a common practice among our survey participants, with introducing variables to explain what a specific value or operation means~\hist{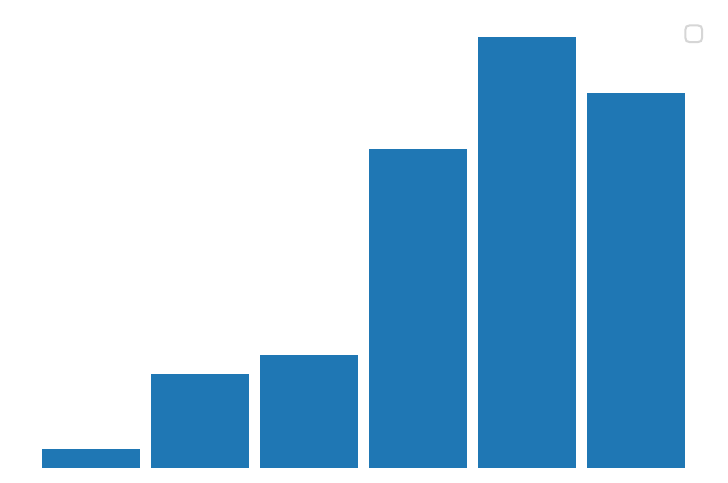}, renaming the test method to better explain what it tests~\hist{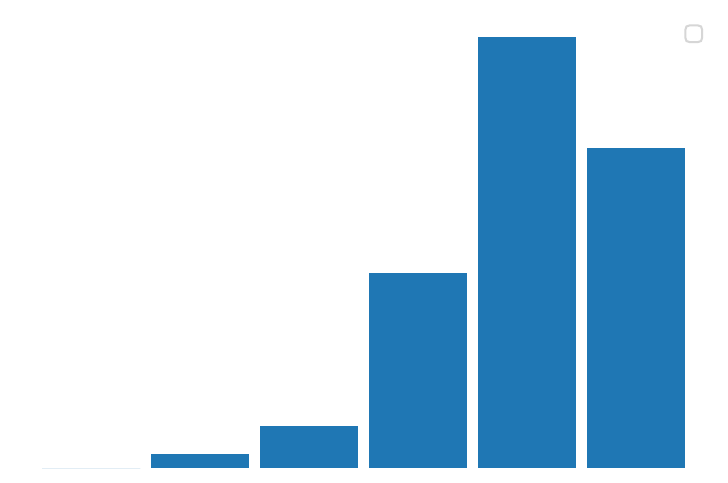}, reorganizing the order in which methods appear in the class~\hist{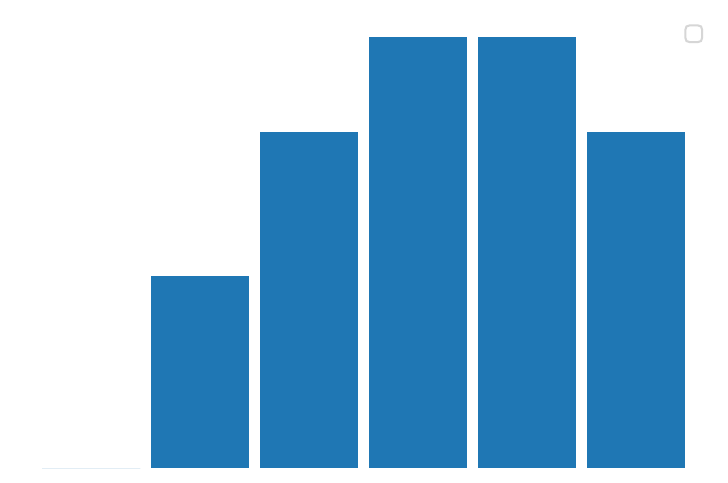}, and removing test duplicates~\hist{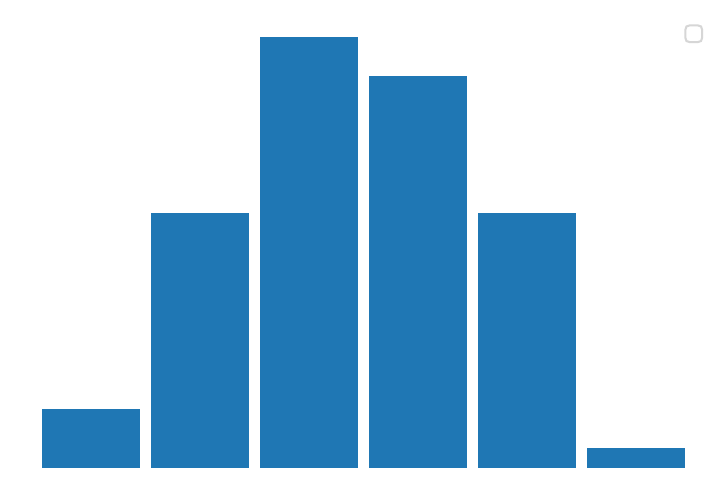} being popular. Also in line with our observations, writing code comments is not so often done by participants~\hist{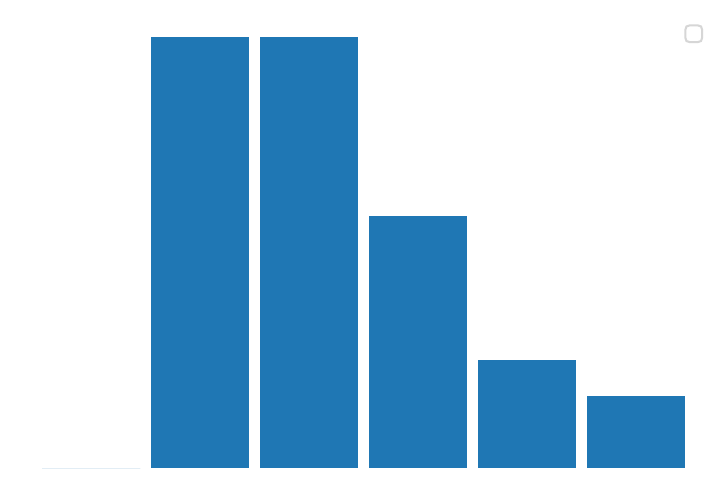}. Interestingly, participants also affirm to write ``buggy test methods''~\hist{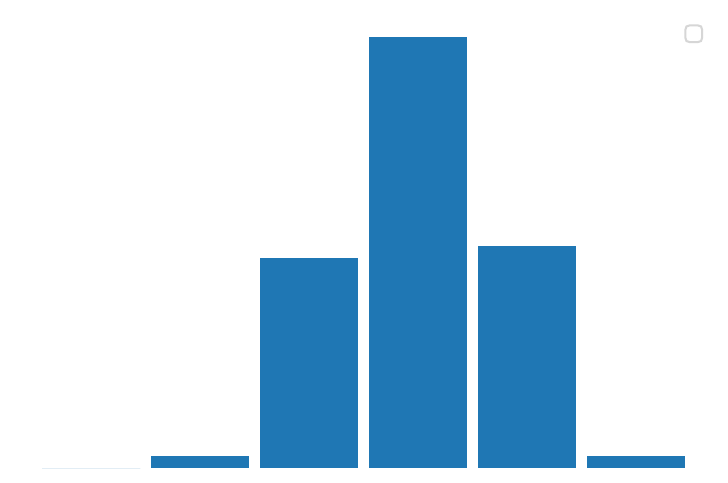}. Although only one person affirms to do it all the time, 25\% affirm to do it frequently, and 48\% to do it occasionally.

\begin{framed}
\noindent Main takeaways (\concept{Test code}):

\begin{enumerate}[label=(C\arabic*),leftmargin=*]
    \item Developers commonly reuse previously written test methods by means of copying and pasting.
    
    \item The input values defined in previous tests often serve as inspiration for the input values of the next tests, although surveyed developers affirm that not to be a regular behaviour.
    
    \item Developers perform several refactoring operations in their test code. Extracting variables (to explain an input value) and renaming the test methods (to explain what the test cases are about) are among the most popular ones.
    
    \item Bugs in test code happen, often due to copying-and-pasting the previous test method or due to the wrong definition of the test assertion.

\end{enumerate}
\end{framed}

\subsection{Adequacy criterion}

Finally, the last concept we discuss is the \concept{adequacy criterion} or, in other words, how participants decided that their developed test suites were ``good enough''. We observed different tactics used by participants.

Six participants (P1, P3, P4, P5, P6, P12) made use of code coverage reports to understand how close they were to being done. All the aforementioned six participants but P4 wrote a few test cases before running the code coverage. P3, for example, just before running the report, said: \textit{``Ok, let me see what I am already covering.''}; P5, near the end of the task, stated: \textit{``I believe I covered everything.''}. He then checked his perception against the code report. P4, on the other hand, systematically checked the code coverage report after every single test case he wrote.  

Exploring the production code (P1, P7, P8, P9, P11, P12) and the test cases produced so far (P4, P7, P9, P10, P11, P12), without the help of code coverage tools, was also a common approach used by the participants near the end of the task. P1, near the end of the task, stated: \textit{``Now, I'm gonna skim the [production] method to see if I missed something.''} P8, after looking at the production code for missing tests, stated: \textit{``I'm ready. I do not see [in the production code] any other test case that I can add.''} Interestingly, from an observational point of view, we did not observe any participants doing a thorough comprehension; rather, most of them skimmed the source and test code quite lightly. 

Finally, only three participants (P1, P3, P5) performed what we call ``documentation coverage analysis''. These participants went back to the documentation and systematically looked for what they were still not testing. P1 even copied the documentation to the test file as to have it closer to the tests. P3 systematically looked sentence-by-sentence of the documentation. 

\highlight{Survey results.} The survey results also show that participants make use of varied sources of information to decide whether they have tested enough. Interestingly, the distribution of answers among the four options we provided was quite similar. Using code coverage, as also mentioned before, is a popular choice with 19\% of participants using it all the time, and 35\% using it frequently~\hist{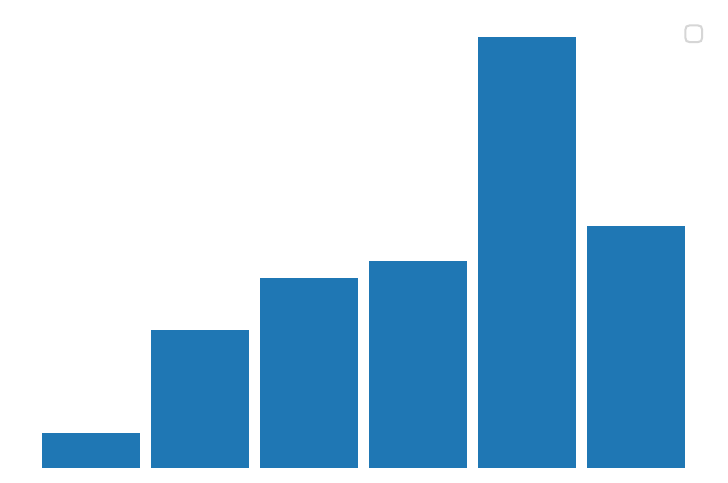}. Manually exploring the source code as a way to ensure everything is tested is performed all the time by 23\% of participants, and frequently by 40\% of them~\hist{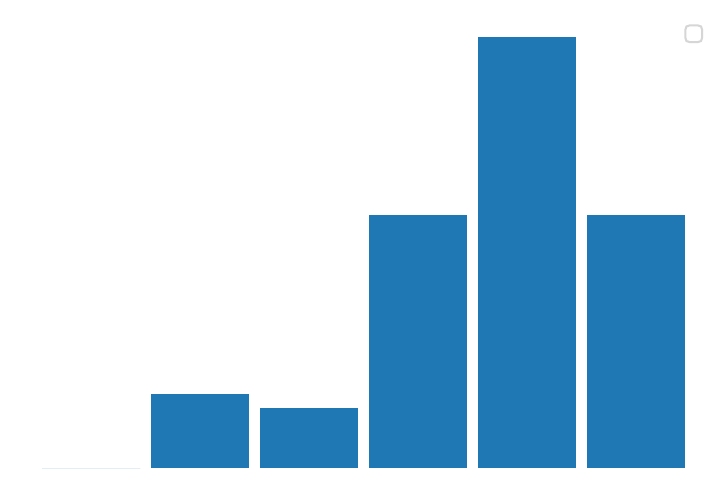}. Re-reading the documentation is done by 14\% of the participants all the time, and by 30\% of them in a frequent manner~\hist{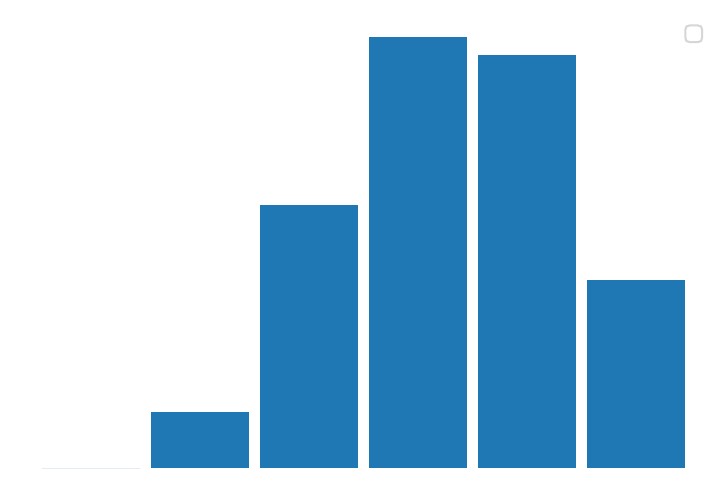}. Finally, participants also may use their personal experiences to decide whether they have tested enough; 10\% affirm to use it all the time, and 37\% of them to use it frequently~\hist{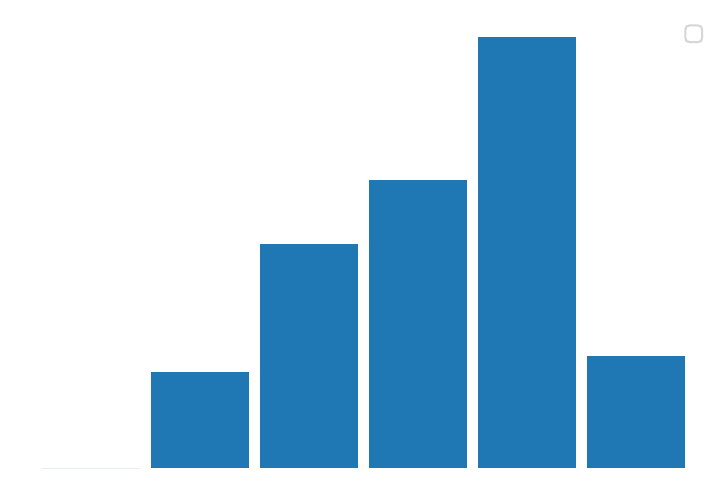}.

\begin{framed}
\noindent Main takeaways (\concept{Adequacy criterion}):

\begin{enumerate}[label=(A\arabic*),leftmargin=*]
    \item Developers use code coverage and revisit the production and test code as a final check to ensure they covered everything.
    
    \item Systematically revisiting the documentation is a less popular adequacy criterion.
    
    \item While not systematically, the personal experience a developer has with software development and testing also plays a role in deciding when to stop testing.

\end{enumerate}
\end{framed}

\section{Testing strategies}
\label{sec:testing-strategies}

The framework we presented above (and illustrated in Figure~\ref{fig:framework}) depicts the main concepts that one can use to explain how developers engineer test cases. As we explained, developers engineer test cases after exploring the documentation, the source code, or using their own mental models of the program's behavior. In this section, we study which \textit{sources of information} developers use and how often.

In the following, we describe the three different so-called strategies that we observed from our participants. We name these strategies after the source of information that developers gave more emphasis to. The three different strategies can also be observed in Table~\ref{tab:events-strategy} as a sequence of events, per participant.

\begin{itemize}
    \item \textbf{Strategy 1: Guided by documentation.} In this strategy, developers intensively rely on the documentation as a main source of information for deriving test cases, and very little on the source code itself. In other words, developers that follow this strategy kept going back to the documentation after creating every test case; the implementation itself was not used much as a source of information. In a way, this strategy is similar to what is known as black-box testing. This strategy was applied by P1, P2, and P9. One participant (P9) even systematically applied domain testing~\cite{kaner2013domain} techniques to derive the test cases. A slight variation of this strategy is to, once test cases are done, leverage the structure of the source code as a way to check whether tests are missing and, in such case, to augment the test suite. This augmentation strategy was applied by P5.
    
    \item \textbf{Strategy 2: Guided by source code.} In this strategy, developers build an initial intuition of the program by reading its documentation, and later are purely guided by the structure of the code (often supported by code coverage tools). More specifically, developers follow ``line-by-line'' or ``branch-by-branch'' in the source code and derive tests for each line or branch. As we illustrate in Table~\ref{tab:events-strategy}, developers that followed this strategy barely looked at the documentation of the program. This strategy was applied by P4, P6, P11, and P12. A slight variation of this strategy is to, once test cases are done, resort back to the documentation and explore whether there is any missing test. This strategy was applied by P3.
    
    \item \textbf{Strategy 3: Ad-hoc (or mixed).} In this strategy, participants leverage the documentation, the source code of the program under test, and the mental model they build out of both sources, to engineer test cases. Participants do not clearly stick with a single source of information as main source, and may resort back to both at any given time. This strategy was applied by P7, P8, P10, and P13.

\end{itemize}

\begin{table}
\caption{The different strategies applied by the participants (N=13). The bars represent the sequence of actions during the analysis. The number in parenthesis indicates the total number of actions for that participant. \tsetwo{Actions related to \textcolor{Turquoise}{\concept{mental model}} are represented as light blue, \textcolor{Salmon}{\concept{documentation}} as red, and \textcolor{RoyalBlue}{\concept{source code}} as dark blue}. The size of the block does not indicate time, but number of actions in a row from that same concept.}
\label{tab:events-strategy}
\renewcommand{\arraystretch}{0} 
\resizebox{\columnwidth}{!}{%
\begin{tabular}{p{0.7cm}l}
\toprule

\multicolumn{2}{p{7cm}}{\textbf{Strategy 1: Guided by documentation}} \vspace{1mm}\\
\raisebox{2mm}{P1 (25)} & \includegraphics[width=0.4\textwidth, height=6mm]{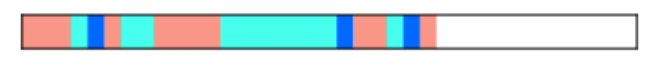} \\
\raisebox{2mm}{P2 (10)} & \includegraphics[width=0.4\textwidth, height=6mm]{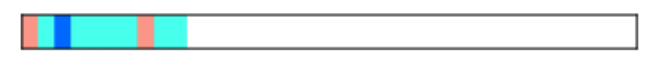} \\
\raisebox{2mm}{P5 (16)} & \includegraphics[width=0.4\textwidth, height=6mm]{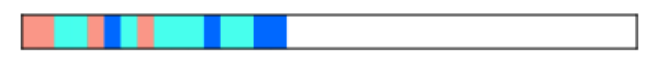} \\
\raisebox{2mm}{P9 (19)} & \includegraphics[width=0.4\textwidth, height=6mm]{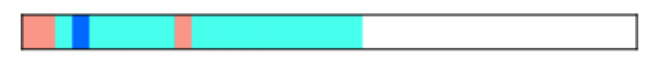} \\
\hdashline

\multicolumn{2}{p{7cm}}{\vspace{1mm}\textbf{Strategy 2: Guided by source code}} \vspace{1mm}\\
\raisebox{2mm}{P3 (15)} & \includegraphics[width=0.4\textwidth, height=6mm]{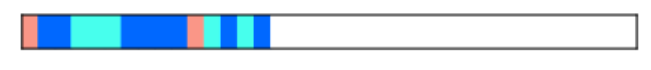} \\
\raisebox{2mm}{P4 (15)} & \includegraphics[width=0.4\textwidth, height=6mm]{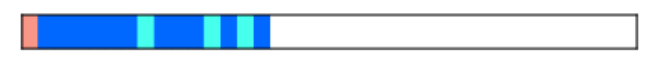} \\
\raisebox{2mm}{P6 (38)} & \includegraphics[width=0.4\textwidth, height=6mm]{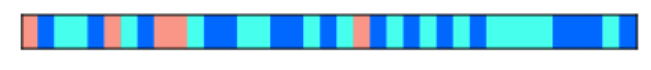} \\
\raisebox{2mm}{P11 (38)} & \includegraphics[width=0.4\textwidth, height=6mm]{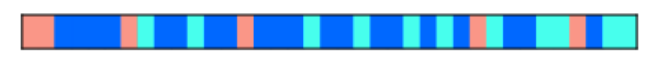} \\
\raisebox{2mm}{P12 (11)} & \includegraphics[width=0.4\textwidth, height=6mm]{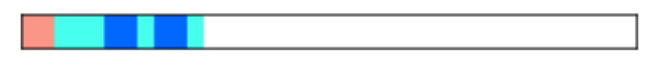} \\
\hdashline

\multicolumn{2}{p{7cm}}{\vspace{1mm}\textbf{Strategy 3: Ad-hoc (or mixed)}} \vspace{1mm}\\
\raisebox{2mm}{P7 (27)} & \includegraphics[width=0.4\textwidth, height=6mm]{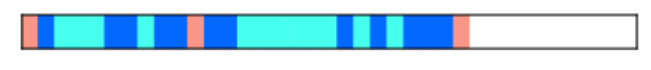} \\
\raisebox{2mm}{P8 (20)} & \includegraphics[width=0.4\textwidth, height=6mm]{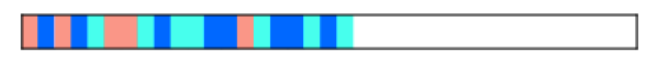} \\
\raisebox{2mm}{P10 (11)} & \includegraphics[width=0.4\textwidth, height=6mm]{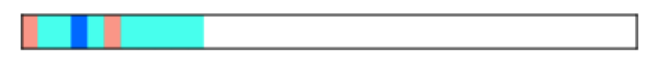} \\
\raisebox{2mm}{P13 (10)} & \includegraphics[width=0.4\textwidth, height=6mm]{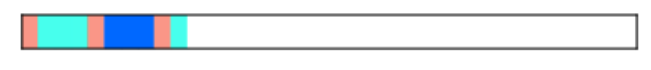} \\
\bottomrule

\end{tabular}
}
\end{table}

\tsetwo{We can see that participants that are guided by the documentation (strategy 1) indeed make little use of the source code (i.e., few to no dark blue events). Most of the tests come from the mental model (light blue events) that they build from the documentation (red events). P5 is the only participant that somewhat pays attention to the source code, but more at the end of the task (the dark blue event in the middle of the task is because P5 made use of the debugger).}

\tsetwo{On the other hand, participants that are guided by source code (strategy 2) intensively rely on it to engineer test cases (many dark blue blocks).} Finally, developers that follow an ad-hoc method (strategy 3) have a more mixed approach, varying between the documentation, source code, and their mental models. Interestingly, we note that all developers, regardless of their strategy, leverage the documentation at the beginning of the task as to build an initial understanding of the problem.

\highlight{Survey results.} The survey results show a clear preference for the ad-hoc method.\footnote{We conjecture that this might be caused by developers not being able to realize whether they tend to focus more on a specific source of information. Future work should explore whether developers have the right perspective about how they leverage different sources to engineer test cases.} Thirty-one participants (or 43\%) claim to explore both the documentation and the source code, whichever is best, to write tests. Moreover, 17 participants (23\%) affirm to be purely guided by the source code, and 14 participants (19\%) to be purely guided by the documentation. Only five participants (6\%) go directly to the test code and explore the behaviour of the method from there.\footnote{We did not list this specific strategy in this section, but the survey question was triggered by our observation that participants often use test code to explore the behaviour of the method.} The remaining five participants opt for an open answer. Two of them mentioned Test-Driven Development (TDD)~\cite{beck2003test}, another participant says that they also communicates with product owners and business teams to understand the expected behaviour, and another says it is a mixture of the options we give in the survey.

When we asked participants to explain their rationale about the strategy they follow, we observe a set of main relevant topics: the lack of good documentation and the reliance on communication with stakeholders to deeply understand the software under development, the presence of good documentation and how much it helps to test, how much documentation and source code complement each other as a reason to use both, the use of previous experience and knowledge of the domain to support more ad-hoc testing processes, and test-driven development.

Participants mention indeed the lack of good documentation (15 out of 72 survey participants). More specifically, participants complain about it being non-existent, lacking clarity, being incomplete or too minimalistic, not discussing corner cases, being ambiguous, and not always being up-to-date with the implementation. As a good example, one participant says: \textit{``In the environment I'm in, the documentation often does not exist and, when there is one, it lacks quality and clarity. That is why going straight to the source code is more common to me.''} Nevertheless, participants seem to find their way by leveraging communication channels with product owners and stakeholders in general (7 participants). One participant even affirms that \textit{``sometimes it is just faster / more practical to go to the analysis or even work with them in pairs; my relationship with them is my main source of understanding of the system, and so documentation works as a secondary source.''} The experience and knowledge that they accumulated over time, as software developers and in the domain of the system, also helps participants in understanding what to test, especially in cases where the documentation is not available (5 participants). 
    
On the other hand, participants that do have access to good documentation in their working environments seem to perceive it positively when it comes to supporting the testing (7 participants). A participant says: \textit{``Today I work together with an architect that documents all features we are developing. Therefore, creating tests is just much easier.''} Another interesting remark came from a participant that identified him/herself as a tester. For them, given that they only focus on testing, their main source of information has to be the documentation.\footnote{We conjecture that this survey participant has no access to the source code and only performs black-box testing.} Another participant also mentions that s/he prefers to focus on the documentation, as it is less biased.

Several participants also affirm that both documentation and source code are important for testing (12 participants). All answers are about how the two sources complement each other (i.e., the documentation explains what the program should do whereas the source code explains how the program does it) and that more information is always beneficial to the testing process. A participant states: \textit{``With the understanding of the documentation and my eyes on the source code, I can more clearly see how the class works and therefore I can test it in the best way possible.''}

Finally, eight participants also mention that they follow a Test-Driven Development approach. Differently from what participants did in the observational study, these participants affirm building the tests together with the feature itself. Interestingly, a participant says TDD gives them more confidence in knowing that they are not doing something wrong, leading them to less mental effort.

\begin{framed}
\noindent Main takeaways (Testing Strategies):

\begin{enumerate}[label=(T\arabic*),leftmargin=*]
    \item Developers mainly use three different strategies: intensively guided by documentation, intensively guided by source code, and ad-hoc. When also developing the feature as part of the process, some participants make use of TDD.
    
    \item Using the other source as a way to validate and augment the engineered test cases (e.g., using source code to augment tests derived from the documentation) is a possible variation for the strategies that tend to focus more on either documentation or source code.
    
    \item While we see a clear distribution of strategies among participants in the observational studies, surveyed developers claim to use a more ad-hoc approach, where both sources are used.
    
    \item The lack of good documentation to support the testing process makes many developers resort more to source code than they want.

\end{enumerate}
\end{framed}

\section{Related Work}

In this section, we discuss the current related work and how their findings relate to the observations we make in this paper. More specifically, we divide this section into two parts. First, we discuss empirical studies that have explored the behaviour of developers when testing their software systems. We then draw parallels between their findings and our observations. Second, we discuss papers that propose tools to support the testing process. In both sections, we propose studies that should be tackled by empirical software engineering researchers in the near future.

\subsection{Developers' behaviours}

\highlight{Focus on testing.} Beller et al.~\cite{beller2015and}, after monitoring 416 software developers closely for around five months, concluded that the majority of the developers do not really test the code they produce (also because a lot of the projects do not contain test suites), that developers rarely run tests in their IDEs, that TDD is not widely practised, and that although developers estimate that they spend half of their time testing, they only spend a quarter of their time. Beller at al.~\cite{BellerTSE2019} later extended the study to a total of 2,443 developers in four different IDEs (for Java and C\#) and observed them for 2.5 years. The same results still held. Interestingly, the participants in our study executed their test cases very often. The difference between our observations and Beller et al.'s studies might be explained by the fact that our participants were fully focused on the testing activity, whereas Beller et al.'s were monitored during all their activities. Similar to our discussion on the need for more systematic testing, we argue that developers having clear testing objectives may help them to focus.
Another key observation from both studies that relates to our study, is that the most typical reaction that developers exhibit when confronted with a failing test is to read the test code and the production code. Understanding exactly what the test code exercises in the production code seems to be a vital task when understanding the cause of the failed test. We conjecture that better tool support would help developers in understanding the cause of the failing test faster.

In a large-scale survey with 5,971 responses, where professional software developers self-reported on their daily activities and reflected about what made their workdays good and typical, Meyer et al. report on ``testing days''~\cite{MeyerTSE2021}. During a \emph{testing day}, developers spend considerablyt more time on testing compared to other activities. Meyer et al. highlight how these testing days happen more frequently among junior developers, that developers mostly consider these testing days to be ``good workdays'', and that ``developers spend more time learning new things [on these days] than on other days.''

Itkonen et al. have explored manual testing practices of 11 software engineering professionals~\cite{itkonenESEM2009}. They found that the observed software engineers use either exploratory testing strategies, or are guided by forms of documentation to guide their testing activities. The important role of documentation that was observed by Itkonen et al. is also apparant from our own observations. 

\highlight{The use of the debugger.} We also observed our participants making intense use of the debugger (takeaway S4). Using the debugger to understand the (dynamic perspective of the) code is also a pattern that has previously been described, as both Demeyer et al.~\cite{OORP} and Spinellis~\cite{effectivedebugging} call this ``stepping through the execution''. Moreover, in their field study, Beller et al.~\cite{bellerICSE2018} observed that around 30\% of the developers use the debugger regularly. Given that the use of the debugger seems to be common, also when writing tests, we suggest researchers in the future to explore how debuggers can be enhanced to support, not only the understanding of the code, but also the creation of test cases.

\highlight{Refactor to understand.} As for rare behaviours, P7 was the only participant that performed slight modifications to the production code to better understand it. While other participants may not have done it due to our experimental settings, surveyed developers also affirm not to do it often. While rare in the context of engineering unit tests, the more general pattern ``refactor to understand'', is a well-described approach to improve program comprehension~\cite{OORP, DuBoisCSMR2005}. We suggest future studies to understand why developers do not apply these comprehension patterns in test code and whether their application would bring benefits to the testing process.

\highlight{Characterisation tests.} We have also seen our participants writing tests as a way to characterise what the code does (takeaways M2 and M3). In other words, our participants often wrote a test that they knew would fail, only to improve the assertion later on when they had complemented their mental model with the information from the failing test. This iterative process resembles a reverse test-driven development strategy. The same behaviour was already observed in the practitioners' literature. Feathers makes a case for writing ``characterisation tests'', or tests that help the developer understand what the code actually does~\cite{workingeffectivelywithlegacycode}. Given that characterisation tests play such a fundamental role in the testing process, we suggest future work to explore best and more productive ways of conducting such activity.

\highlight{Bad weather testing.} We have observed that developers do not solely focus on ``happy path'' test cases, but also spend time on engineering boundary and exceptional test cases, the so-called ``bad weather tests'' (takeaway TC2). The importance of these bad weather tests has been discussed in the literature before by {\c C}alıklı and Bener~\cite{calikliSQJ2013}, Salman et al.~\cite{salmanEMSE2019}, and Teasley et al.~\cite{Teasley1994}. In particular, {\c C}alıklı and Bener~\cite{calikliSQJ2013} highlight how confirmation bias often drives software engineers to focus on testing positive scenarios, thereby neglecting bad weather tests. Salman et al. also investigated whether confirmation bias can be linked to time pressure when writing tests, noting that confirmation bias can be observed regardless of time pressure~\cite{salmanEMSE2019}. This was not the case for our participants, as all of them focused on bad weather tests. Again, we conjecture that such a focus could be caused by the fact that they were focused on the testing activity. 

\highlight{Test code refactoring.} The participants in our study often refactored their test code (takeaway C3). While we have a large body of literature on test code quality and how to refactor test code (e.g.,~\cite{van2002video, van2001refactoring, meszaros2007xunit}), test smells~\cite{palombaICSME2018,spadiniICSME2018} are still quite prominent in test suites. For example, Bavota et al.~\cite{bavota2015test}, after manually analysing around 1,000 test classes, noticed that 86\% of the JUnit tests exhibited at least one test smell. Tufano et al.~\cite{tufano2016empirical}, after analysing the life cycle of test smells, observed that test smells tend to be introduced right in the very first version of the class. Moreover, the test smells tend to stay in the codebase for a long time. 

We did not observe any severe test smells~\cite{meszaros2007xunit} in the code produced by the participants during the observational study. Our participants may have paid more attention to the code quality because they were being observed, a common threat in observational studies. Regardless of that, the fact that participants refactored their test code quite often shows that they indeed care about code quality. Caring about test code quality has also been observed by other researchers, e.g., Spadini et al.~\cite{spadini2018testing} noticed that most comments in the code reviews of test code are related to improving the code. Similarly, Daka and Fraser~\cite{DakaISSRE2014} have observed that maintainability of the test cases is a primary concern among the respondents.

Prior studies have also highlighted the intertwined nature of refactoring and testing. In particular, Moonen et al.~\cite{DBLP:series/springer/MoonenDZB08} have shown how around 1/3 of the refactorings from Fowler~\cite{Fowler1999} somehow invalidate unit tests, making the refactoring of test code a necessity. In this context, they have coined the term ``test-driven refactoring'', where the tests are first refactored, and then the production code is adjusted to fit.
Kashiwa et al. build on this investigation and mine software repositories to find refactorings that break the tests~\cite{KashiwaICSME2021}. They observe that in practice 
most types of refactoring operations do not break test suites. In case the refactoring does break a test, fixes are typically small. However, some refactoring operations, like \emph{add paramater} inherenty break the test. 
In other research, Vonken and Zaidman have studied how useful the presence of test code is in the light of production code refactoring. Their study indicates that students who dispose of tests during refactoring of production code perform the refactorings both more quickly and more correctly~\cite{DBLP:conf/wcre/VonkenZ12}. 

\highlight{Test-Driven Development.} Test-driven development (TDD) is an agile software development approach in which software developed in small iterative testing-coding cycles~\cite{SantosEMSE2021}. In a family of experiments, Santos et al. investigate why the results of several studies revolving around TDD have provided a set of mixed observations when it comes to perceived benefits of TDD. They argue that the lengthy of the observation, the unit of analysis, the project length, and the programming environment all contribute to the outcome of the studies. In related work, Beller et al. tracked 2443 software engineers in the wild, and noted that TDD is not frequently applied~\cite{BellerTSE2019}.

Zaidman et al. found that test code does not always immediately co-evolve with production code~\cite{DBLP:conf/icst/ZaidmanRDD08,zaidmanEMSE2011,marsavinaSCAM2014}. In fact, they have observed intense periods of working on production code, after which developers transitioned into test-centric periods of development work. Contrasting the test-driven development idea, this phenomenon corresponds more to the notion of test-last development~\cite{fucciTSE2017}, albeit sometimes with longer time intervals. Fucci et al.~\cite{fucciTSE2017} investigated whether a test-first or test-last development strategy actually influences the external quality of a software system. They concluded that it is not so much the test-first or test-last strategy that influences external quality, but rather a short development cycle that positively influences external quality.

\highlight{Experience and prior knowledge in testing.} In the paper by Yu et al.~\cite{yu2019comprehending}, the authors show that experience and prior knowledge may play a role in the developers' productivity (i.e., knowledge of domain reduces test read time) and quality of tests (i.e., more experience with testing and development influence the developers' abilities to produce more test cases). 

The 13 developers as well as the 72 surveyed developers have varied levels of experience. While we did not objectively measure the influence of experience, we indeed observed some developers being more productive than others. P12 (14 years of experience as a developer, 10 years of experience with testing), for example, finished the task within only 12 minutes. P4, who has the exact same number of years of experience as P12, finished the experiment in 32 minutes. On the other hand, P9, also quite experienced (12 years of experience as a developer, and 6 with testing), took a little more than 1 hour. Interestingly, P9 also have affirmed to have previous knowledge regarding the method he was supposed to test. Finally, we also note that experience may play a role in deciding when to stop testing (takeaway A3). Future work should focus on understanding what makes some developers more productive than others.

Moreover, it is interesting to note that, although we categorise the different strategies into three buckets (guided by documentation, guided by source code, and ad-hoc), we see great variability in how they are applied, even among developers that opt for the same strategy. 
While the main goal of this study was not to compare the different strategies in terms of performance and achieved coverage (the 13 data points we have also do not enable us to generalise), from the numbers in Table~\ref{tab:demographics-full}, we see that all participants were able to achieve similar coverage.
Nevertheless, understanding whether one practice may lead to a comprehensive test suite faster than other practices might be useful information for developers. In the program comprehension field, evidence suggests that different developers apply different strategies, such as top-down or bottom-up comprehension~\cite{detienne2001software}, and that developers can be equally productive in both. We see similar studies focused on testing as necessary.

Itkonen et al.~\cite{ItkonenTSE2013} have investigated prior knowledge in the context of \emph{exploratory testing} and have reported that software testers apply knowledge of the system under test and its application domain, including users’ needs and goals. More specifically, personal knowledge is applied  in exploratory testing to evaluating the overall behavior of the system,
comparing the features with other features (of similar systems), and applying knowledge of earlier versions of the system.

Bai et al. report on perceptions of students on testing~\cite{baiITICSE2021}. In particular, students report that they encounter the following two key challenges: (1) understanding the source code implementation to test, and (2) understanding when to step testing.

\highlight{Mocking.} Mocking is a common technique used by developers when they face more complicated pieces of code to test. In particular, Spadini et al.~\cite{spadini2017mock, spadini2019mock} explored how different Java systems make use of mocking. The authors observe that developers tend to mock infrastructure classes (e.g., classes that access databases or webservices) and/or classes that are too complex to be instantiated in the test code. Moreover, authors observe, while mocks may facilitate testing, the test code becomes more coupled to production code. In fact, the authors have observed a significant number of times where a change in the production code required a change in the code of the mock. We argue that mocking brings a whole new dimension to the testing process, as developers have to decide whether a dependency should be mocked, to reason about the contract of the dependency being mocked, as well as to define the behaviour of the mocked object. While the work of Spadini et al.~already shed some light on the decision-making process of developers, this data was collected by means of interviews. We suggest future observational studies to explore the challenges that developers face when reasoning about the use of mocks.

\highlight{Unit vs integration testing.} In this study, participants solely focus on engineering unit tests~\cite{DBLP:journals/software/Runeson06,whittaker2000}. While the difference between unit and integration testing might be somewhat blurry in practice (see Trautsch et al.~\cite{trautsch2020unit}), we conjecture that the developers' needs and decision-making processes might be different between the different test levels. Therefore, we argue that future studies should investigate how developers engineer integration and system tests. However, based on the observations of Greiler et al., we do know that unit testing plays a key role in open source projects, with unit test suites comprising thousands of test cases. System, integration, and acceptance testing, on the other hand, are adopted and automated less frequently~\cite{greilerICSE2012}.

\highlight{People analytics and the social side of testing.} Finally, people analytics~\cite{singer2015people} has also been explored by the community as a way to engage developers in testing. Pham et al.~\cite{pham2015communicating} proposed an extension to the IDE that shows how much other developers in the team were testing. After an experiment with bachelor students, authors observed that students that were reminded of their lagging test progress were often induced to test more. Moreover, the ``feeling of competition'' also made students want to write more tests. 
In another study by Pham et al.~\cite{pham2013creating}, after interviewing and surveying active open-source developers, the authors reported other social factors that may affect the testing behaviour of developers, such as the current existence of test suites in the project or how hard the initial barrier to provide test cases is in the project.
There was no way for our participants to exercise their ``social side'' in our experiment, given that participants worked alone on their tasks. An interesting future work would be to repeat our observational study, but inside of different software companies, where different social factors may also play a role in how developers test.

\highlight{Cognitive aspects.} 
As software testing is an  intellectual activity that requires analysis, reasoning, decision making, abstraction and collaboration, Enoiu et al. present a theory on software testers' cognitive proceses~\cite{enoiuQRSC2020}. From a ensuing study, the authors found that, on average, 39\% of their time was spent on analyzing their knowledge regarding the test goal and planning different approaches on how to create test cases.  Participants organized information via inferencing and case-based reasoning. When they had gathered the necessary information, the subjects to create test scripts.

\highlight{Best practices.} Kochar et al. have interviewed 21 and surveyed 261 practitioners with regard to their best practices when it comes to developer testing~\cite{KocharICSE2019}. They have established 29 characteristics of good test cases in 6 dimensions (i.e., test case contents, size and complexity, coverage, maintainability, bug detection and others). Some of the observations from Kochar et al. reinforce our own observations, for example: (1) their observation on the importance of traceability links between test cases, code, and requirements, (2) that code coverage can be used to steer test engineering activities, and (3) that the understandability of the test cases is important.

In other work, Athanasiou et al. determined whether the code quality of test cases influencese the speed by which developers can react to maintenance operations~\cite{AthanasiouTSE2014}. Their findings highlight how developers are quicker to implement code additions if they have high-quality test to protect them against regressions.

\subsection{Tool support}

\highlight{Code coverage.} Berner et al.~\cite{berner2007enhancing} report an experience with a code coverage tool in a project at the Swiss National Bank. After introducing a tool that would enable developers to follow the coverage of their systems over time, authors made interesting observations regarding how the behaviour of developers changed. For the senior developers, the authors note that the coverage rate was stalled; however, after the introduction, the coverage rate increased immediately, but moderately, and slowed down again a month later. For the junior developers, while they required a longer period to get comfortable with the visualizations, the duration of the effect was longer and the relative increase of the coverage rate was higher when compared to the senior developers. 
For both junior and senior developers, the authors observed an increase in the number of tests focused on error handling.
Lawrance et al.~\cite{lawrence2005well}, on the other hand, in a controlled experiment with 30 experienced software developers, did not observe significant differences between developers using and not using code coverage visualization tools, both in terms of test effectiveness and amount of tests written. In addition, the authors observed that even more experienced developers may perform counterproductive testing strategies, such as changing the parameters of a method under test or deleting failed tests.
In our study, we observed some developers making use of the code coverage tool support provided by Eclipse and IntelliJ to understand what they had already tested and what to test next (takeaways S5 and A1). For the developers that opted for such tools, we perceived the coverage information to be fundamental in their testing strategy. Because we have mixed evidence supporting the use of code coverage visualizations, we suggest more replications of such studies.

Also related to code coverage, we note that our study focuses on testing a single snippet of code that had no prior tests. Developers also have to evolve a piece of code to, e.g., add new functionality. Elbaum et al.~\cite{elbaum2001impact} have showed that even slight modifications to the production code might drastically affect the coverage of the existing test suite. Hilton et al.~\cite{HiltonASE2018} have studied the evolution of code coverage of a large number of projects, and have established that at the project-level, fluctuations in coverage are hard to discern. As such, they make a case for establishing code coverage at the patch level. Hurdugaci and Zaidman~\cite{hurdugaci2012aiding} have proposed a tool that helps developers in identifying which test cases they should change, given a change in the production code. We suggest future studies to explore how developers behave when extending existing functionality and its related test suite, as their needs might be different when compared to the ones observed in this study.

\highlight{Test frameworks.} Kochar et al. have investigated the test automation culture among app developers~\cite{KocharICST2015}. They found that Android app developers prefer using standard frameworks like JUnit, but that many developers also still like to test manually without making use of any testing framework or tool.

\highlight{Test-to-production traceability.} Regarding the traceability between production and test code, we conjecture that such information may support developers in better understanding the cause of failing tests (takeaways D2, S4, M2, TC1). Other studies have also shown that developers indeed see such traceability as a feature that would help them in testing (e.g.,~\cite{spadini2018testing,prado2018towards}). We see some work in the literature that aims at establishing such links. For example, the work of Van Rompaey and Demeyer~\cite{van2009establishing} evaluated different strategies and showed that a simple strategy such as relying on test conventions already yields highly accurate results. We suggest future user studies to explore how developers would make use of such tools and what benefits they would concretely bring to the process.

\highlight{Test amplification.} The field of test amplification~\cite{DBLP:journals/jss/DanglotVYZMB19,DBLP:journals/corr/abs-2108-12249} (i.e., amplifying the test suite by leveraging the already existing developer-written test cases) is also emerging quite rapidly in the community. In a literature review, Monperrus et al.~\cite{monperrus2017emerging} show that test amplification improves test suites in terms of coverage, mutation score, fault detection capability, and debug effectiveness. A related observation from our study is that, although developers did not make use of any test amplification tools, they seem to ``manually perform test amplification'' quite often (takeaway C2). In other words, developers often make use of test cases they previously engineered as an inspiration for the next test case. 
Also related to help developers in augmenting their test suites, we note the UnitPlus tool, work of Song et al.~\cite{song2007unitplus}. UnitPlus observes the source code being tested and suggests assertions to developers, based on static analysis of the code being tested (e.g., if a method modifies a specific attribute of a class, UnitPlus will recommend for that field to be asserted).
We suggest future observational studies to understand how developers would combine their developer tests with the tests amplified by a tool.

\highlight{Parameterized testing.} We highlight that we did not observe developers making much use of parameterized tests, but we argue that the type of tests they have written is a perfect fit for such a feature. Researchers have explored the possibility of automating the refactoring of test code towards parameterized tests. While some work has been done in automatically generating parameterized tests (e.g., \cite{tillmann2006unit,xie2010future}) and retrofitting unit tests to a more generic and parameterized unit testing (e.g., \cite{thummalapenta2011retrofitting}), such works are focused on helping developers in identifying more test cases. 
We are not aware of any work that refactors the existing test code of the developer in a form of a JUnit parameterized test, simply for the purpose of code maintenance, which we suggest as an interesting tool to be developed.

\section{Recommendations}

In this section, we discuss how our observations may be actionably leveraged by toolmakers, software developers, educators, and researchers.

\subsection{Recommendation to toolmakers}

We propose four recommendations to toolmakers:

\highlight{Mechanism to easily derive test skeletons.} 
Developers seem to strongly rely on copying-and-pasting mechanisms when writing test code (takeaway C1). This is clearly understandable given that test methods, especially for programs such as the ones we selected as tasks where they receive a clear input and return a clear output without the need for instantiating other objects or mocks, tend to share a large amount of similar code. JUnit does offer the \texttt{ParameterizedTest} functionality that enables developers to write the skeleton of the test method and passing several inputs via parameters.\footnote{See documentation: \url{https://junit.org/junit5/docs/current/user-guide/#writing-tests-parameterized-tests}.} However, only P1, P5, and P11 reflected on its usage. 
We suggest a tool that notices such patterns in test code (i.e., test methods that are similar in structure, but with different inputs and outputs) and suggests (or even automatically applies the refactoring) to condensate the tests into a single parameterized test.

\highlight{Lightweight and fast code coverage.}
Code coverage seems to be commonly used by many developers (takeaways S5 and A1). However, for that to happen, developers need to explicitly go for the ``coverage'' option in their IDEs. The IDE then commonly runs the entire test suite and presents the results to developers. We argue that such a feedback loop should happen faster (i.e., without the need of running the entire test suite every time), in a less intrusive way (i.e., without really requiring developers to manually select the ``run coverage'' option), and with explicit differences between the current and the previous runs. Understanding what lines of code a single test covers may also bring benefits to developers, as this is often what they do whenever they face a failing test case.\footnote{Somewhat similar to what has been proposed by Tasktop in the past: \url{https://www.tasktop.com/blog/incremental-code-coverage-debugging-tool/}} One possible way of implementing such a tool would be with continuous code coverage calculations (maybe only relying on the currently added or modified test method) in the background. 

\highlight{Detection of ``accidentally buggy'' test methods.} We have seen developers accidentally writing bugs in their test code (takeaway C4). From our observations, we note that most of the bugs were caused by copying-and-pasting, e.g., not changing the inputs or the assertions of the copied test. 
We there argue that tool support is needed in order to support developers in identifying such buggy test methods before they confuse the developer. 

\highlight{Support for testing the most common corner and edge cases.} Testing extreme, corner, and edge cases is a fundamental part of the developers' process (takeaway TC2). 
As a way to support developers in testing their code in a more systematic way, we envision a tool that suggests test cases to developers that they did not write yet. For example, the tool can remind the developer to try out an empty string for each input that is of type string, or zeroes and negative numbers for integer inputs. Moreover, the tool could rely on the structure of the code under test and suggest developers to engineer test cases that force loops to be executed zero or one time only (i.e., similar to what is suggested by the loop adequacy criteria~\cite{howden,zhu1997software}).

\subsection{Recommendation to developers}

We propose five recommendations to developers. Interestingly, these recommendations are not fully novel; many of them are already (\emph{intuitively}) known by our scientific community, e.g., the importance of systematic testing. Our findings show, however, that there is a disconnect between what the scientific community suggests and what developers do in practice. We hope this paper will serve as another piece of evidence supporting the benefits of such practices.

\highlight{More systematic software testing.}
A large chunk of the software testing literature focuses on making the testing process somewhat systematic. For example, books on domain testing (e.g., Kaner et al.~\cite{kaner2013domain}) provide developers with clear sets of steps on how to systematically derive tests after a given requirement, e.g., first identify and characterize the input variables, then determine their type and scale, understand how they are related to each other and how the program uses each of them, etc.; boundary testing techniques (e.g.,~\cite{jeng1994simplified, reid1997empirical, hoffman1999boundary, legeard2002automated, samuel2005boundary}) support developers in ensuring that the boundaries of every partition in their domain input are exercised.

While we categorize the different strategies we observed in three buckets, rarely did we see a developer following a systematic approach within that strategy (takeaways T1 and T3), with the exception being P9, who followed a similar domain testing strategy. Developers that were guided by the source code tended to derive tests following the order in which the code appeared in the program under test, although this was more prevalent at the beginning of the method (where the programs we selected perform pre-condition checks such as inputs not being null), but more ad-hoc later when the ``real implementation'' of the program started. We make a similar observation for boundary testing. While developers did focus on exercising different branches, they rarely focused on systematic boundary testing.

Although most participants achieved a branch coverage of 90\% or more (see Table~\ref{tab:demographics-full}), showing that they can achieve high coverage regardless of their strategy, we argue that the systematic application of such techniques would help participants in covering more cases. For example, 
P12, a highly experienced software developer (14 years of experience) and tester (10 years of experience) achieved 100\% branch coverage in under 13 minutes. He was clearly the fastest participant in finishing the task. However, we see that he achieved ``only'' 88\% mutation coverage. When looking at which mutants survived, we see that two ``conditional boundary changed'' mutants in conditions \texttt{if (pads <= 0)} (with condition replaced by \texttt{<})  and \texttt{if (pads < padLen)} (with condition replaced by \texttt{<=}) survived. While a more in-depth understanding of the mutations revealed that the mutant in the second condition cannot be killed (due to an \texttt{if} instruction just before the targeted condition), the first mutant can be actually killed. In fact, we observe the same in P4, P6, P9, and P11, as all of them achieved 100\% code coverage but missed more thorough testing in that boundary. None of these developers has noticed missing such tests during their tasks. 

\highlight{Do not make assumptions about the behaviour of the program.} 
We have seen seven cases where the developer simply accepted that the behaviour of the method was correct without really double-checking with the documentation (takeaway S3). This can be a consequence of the tasks we have chosen for this experiment; the tasks are all about common string manipulations that developers have (and use) in libraries, and ``guessing'' what they have to do in such cases is possible. Nevertheless, we saw cases where developers were clearly in doubt about the expected behaviour and did not resort back to the official source to consult. Interestingly, in the survey, developers affirm to not trust the outcomes of a method as the correct output. Still, we suggest developers to ``fight their instincts'' and, whenever in doubt, resort back to the documentation and/or a trusted source about the expected behaviour of the program.

\highlight{Have a clear adequacy criterion.} 
Several participants did not have a clear adequacy criterion to decide when they were done testing (takeaways A1, A2, A3). Many relied on their experiences while skimming the source code and the documentation one last time. This means we observed uncertainty from most participants at the end of the experiment: \textit{``I guess I am done''} and \textit{``I think I tested it all''} were common final sentences. While this again could be a consequence of our study being quite a controlled one, and developers may have an adequacy criterion defined as a team, we recommend developers to define a clear adequacy criterion to avoid such uncertainty.

\highlight{Vary testing inputs.} 
We observe developers often using, maybe unintentionally, the same base input for all the tests, e.g., a string \texttt{abc} that becomes \texttt{abd} in one next test and \texttt{abcc} in another one (takeaway C2). We conjecture that such small variations in the same input help developers in building knowledge about the behaviour of the program faster, and maybe that is why several of them do it. Interestingly, survey developers seem not to make use (or realize that they make use) of a base seed as often as we observed in the observational study. Nevertheless, varying test inputs is important~\cite{whittaker2009exploratory}. We therefore recommend developers to keep using base inputs to build understanding, and to complement such test cases with varying inputs once they fully understand the behaviour of the program. 

\highlight{Ensure good documentation to support the testing process.} 
We have observed developers trying to resort back to the documentation and not finding the precise information they needed (takeaways D1, D3). The surveyed developers also seem to perceive this issue as a recurrent one in their daily jobs. While somewhat out of the scope of a testing process, we argue that software development teams should invest their time in either building up documentation that can be used by teams to improve their testing or, as agile methods suggest, to have an active product owner that developers can resort back to.

\subsection{Recommendation to educators}

We propose two recommendations to educators:

\highlight{Make developers aware of the framework and the benefits of the different sources of information.} We argue that understanding and being aware of how developers reason about testing is a fundamental stepping stone towards improving their behaviour. First, we recommend educators to explain to novice developers the basis of the framework we propose in this paper. We hope that, with that knowledge, developers will better understand how test cases are engineered, how documentation, source code and the mental model are used to derive test cases, and how adequacy criteria are chosen.

Moreover, our study shows that different developers seem to give different emphasis to the two different possible sources of information for testing we provided them; some developers made more use of documentation, others made more use of the source code. We note that documentation and source code bring different types of information to the testing process. One would therefore expect developers to make use of both when they are available. Interestingly, in the survey, developers showed a clear preference for an ad-hoc strategy, where both sources are used; we did not see that in the observational study. We conjecture a reason for such a difference would be that, given the familiarity of the participants with the proposed methods to be tested, they did not need to resort too much to either the documentation or the source code. Nevertheless, we recommend educators to explain to novice developers what the different sources of information bring to the table.


\highlight{Teach JUnit best practices, how to use code coverage, and boundary testing.} Moreover, our results also show that developers may benefit from better understanding JUnit (and, e.g., parameterized tests), how to use code coverage tools (not only in continuous integration~\cite{vassalloICSME2016,vassalloICSME2017,bellerMSR2017} but as a tool that supports coding), and boundary testing (so that developers exercise all the boundaries of the implementation more systematically).

\subsection{Recommendation to researchers}
\label{sec:recommendations-researchers}

We propose three suggestions for empirical software engineering researchers.

\highlight{Study how test cases are engineered in different domains, contexts, and software systems with different levels of complexity.} First, while we believe that the complexity of the methods in our study is good enough for us to observe and draw sound conclusions on how test cases are engineered, there are many other cases that are worth investigating. For example, (i) our study does not capture how developers reflect whenever they are testing a class that depends on another class; how do developers engineer test cases for classes that depend on other parts of the system? (ii) testing for specific type of applications, e.g., mobile; how do developers engineer test cases for mobile applications? (iii) how would the developers' behaviors be affected when they are the ones also implementing the production code?, (iv) how do human aspects of software engineering influence how test cases are engineered, e.g., teams with multiple developers, the relationship with the customer of the system, and even tight deadlines, (v) how much does the accumulated knowledge about the software system that developers obtain over time, i.e., the long-term mental model, influence the test case engineering process?, (vi) how much does the level of experience influence the test case engineering process; would novice developers behave differently than senior developers, and if so, how? 

Engineering test cases is a complex phenomenon, highly-dependent on too many factors. Future work should explore the aforementioned points by means of not only observational studies like the one we conducted in this paper, but also through controlled experiments, action research, and ethnographical studies. We argue that the framework we propose in this paper can serve as basis for these future studies.

\highlight{Better understand some of the behaviors we observed.} In this paper, we did our best to report and explain all the behavior we could observe and for which we had data to back up our observations. Nevertheless, there is still more to dive into. For example, we have seen cases where developers misunderstood the documentation, i.e., the documentation is correct but the engineered test case was wrong. How often do such situations happen in real life? How long does it take for the developer to notice their mistake? We have also seen some developers exploring more boundary situations than others; what makes some developers do it, and others not?

\highlight{Understand why the survey participants disagreed with some of our observations.} In a few cases, the survey participants did not agree with what we observed. In other words, they affirm not to do the behavior we saw during the observational study. For example, we noticed that developers tend to jump right into testing even without fully understanding the program under test. Survey participants affirm not to do that so often. We also observed developers engineering test cases out of the mental model that they constructed of the program (through reading the documentation and exploring the source code). In the survey, participants affirm to use the mental model less often than the other methods. While we have no explanation derived from our data, we conjecture that it might be hard for developers to perceive their own behavior when it comes to these two points; after all, it does not make sense to start testing something one does not fully understand although, as we have seen, it does happen. Exploring more about how developers build and use their mental models to engineer test cases is an interesting future work.

Another point of disagreement was in how developers picked specific inputs to the test cases. From our observational study, we could not see any pattern or reason for the specific values that developers chose; we therefore argued that these values do seem to be picked at random. We also observed that, once the first input values are chosen, the next inputs tend to be derived from the first ones. The survey participants affirmed not to select values at random or to use the same seed for the different inputs that often. The developers' lack of awareness on how they pick input values may explain such differences. Another possibility would be that developers do tend to pick values that matter to their business (not at random), which did not happen in our study given that participants were testing utility methods. Future work may focus on understanding how developers pick input values for their test cases.

\section{Threats to Validity}

In this section, we discuss the possible threats to the validity of this study and the actions we took to mitigate them. \tsetwo{We categorize our study as research comprising semi-structured or open-ended interviews, according to the empirical standards for software engineering research~\cite{ralph2020empirical}.}

\subsection{Construct validity} Construct validity is the degree to which our instruments measure what they claim to be measuring. In the case of this study, threats to the construct validity are related to the tasks we provided to participants, to the thinking-aloud protocol, and to the survey we devised.

First, as we discussed together with the recommendations for researchers, the methods that participants tested have limited complexity, and their domains are somewhat familiar for software developers. That being said, we did not observe any unexpected behaviour from participants (e.g., a participant that did not have to look at the documentation or the source code to start doing the task); on the contrary, all participants had their own challenges while testing the programs. We, therefore, argue that the choice of the tasks enabled us to observe real-life behaviour. That being said, as we discussed in Section~\ref{sec:recommendations-researchers}, testing is a complex phenomenon and should be studied from many different angles and contexts, which we leave as future work.

Moreover, developers were asked to record a video and to talk aloud. A first possible threat is whether participants are actually able to think-aloud while performing the task. We note that all 13 participants were very talkative, and that their thoughts were augmented by the video recording. Therefore, we do not have a reason to believe that participants were not able to express their thoughts while testing. Participants also clearly knew that they would be watched later, thus making them susceptible to the Hawthorne effect (i.e., individuals modifying their behaviour just because they know they are being watched) and social desirability bias (i.e., individuals perform in the way society expects them to behave in a specific situation). Because of that, some participants may have been more thorough in the testing process during the experiment than what they really are in their daily jobs, e.g., they tested more bad weather tests than they would. 
Nevertheless, given that our goal was to understand how they reason (and not how thorough they are), we do not see such a change in behaviour as a significant threat. Finally, in the task description, we suggested participants to use Zoom\footnote{\url{https://www.zoom.us}} to record their videos. Although Zoom is a conference tool, it enables people to record their screens without a lot of effort and platform-independent. Zoom, however, gives only 40 minutes to free users. Some participants may have felt pressured to finish their videos within the time limit. We did not ask participants how they recorded their videos or whether that limit played any role. From the videos themselves, we did not notice any participant ``rushing'' to get their testing done and no participant said anything in the videos about them running out of time.

Finally, the survey challenges the main observations in the observational study. For each of the observations, we proposed a statement that could be answered by means of a Likert scale that captures frequency (i.e., how often that particular event happens in a participant's daily job as a developer). Even though we did our best to disambiguate all the statements, they are always subject to the interpretation of the respondents. Before sending to participants, the survey was reviewed by the second author of this paper (using Google Translate's functionality) and by a PhD candidate that has the native language of the survey as first language. We also did not receive any questions about the survey itself from participants. We also make all the questions and answers available in our online appendix for inspection~\cite{appendix}. Moreover, another instance of a possible interference of such social biases is in the survey questions that tackle ``bad behaviour''. For example, we ask how often participants test a method without fully understanding it. Responses there were mostly negative, contradicting what we noticed in the observational study. More replications with different control measures need to be done, to especially understand how often developers resort to ``bad behaviours'', even when they are not willing to admit.

\subsection{Internal validity} Internal validity is the extent to which the evidence we bring supports the claims of this study. In the scope of this paper, possible threats to the validity of this work stem from the qualitative analysis we conducted from observing the 13 participants in action.

\begin{figure}
\centering
\begin{footnotesize}
\begin{tikzpicture}
\begin{axis}[visualization depends on=rawx \as \myx,
    align=center,
    title = {Number of unique codes},
    y=0.075cm,
    symbolic x coords={P1,P2,P3,P4,P5,P6,P7,P8,P9,P10,P11,P12,P13},
    axis line style={opacity=0},
    major tick style={draw=none},
    ytick={0,25,50},
    xtick=data,
    enlarge x limits={abs=0.25cm},
    ymin = 0,
]
\addplot[ybar,fill=gray,draw=none] coordinates {
(P1,22)
(P2,27)
(P3,30)
(P4,31)
(P5,32)
(P6,34)
(P7,40)
(P8,43)
(P9,48)
(P10,50)
(P11,51)
(P12,52)
(P13,54)
};
\end{axis}
\end{tikzpicture}
\end{footnotesize}
\caption{\tsetwo{Number of unique codes after analyzing each participant's video.}}
\label{fig:saturation}
\end{figure}
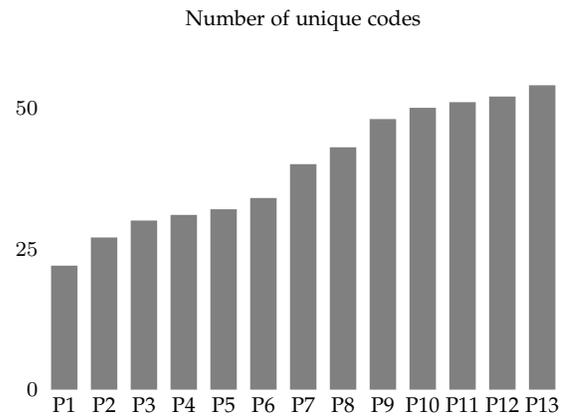

First, the participants were recruited via a single tweet made by the first author. The first author has around 8k followers on Twitter and many of his followers are Brazilian developers. Although our initial goal was to search for participants in rounds and leverage the network of all the three authors, we achieved saturation without the need of a second round. \tsetwo{As shown in Figure~\ref{fig:saturation}, the total number of unique codes in our data only increased by four with the last three participants, while we had obtained 50 such unique codes from the analysis of the first ten participants.} Nevertheless, the participants (followers of the first author who tweets often about software testing) may have biases that we do not know. While we did not observe any particular behavior, we suggest researchers to replicate this work in different communities.

Most of the analysis was conducted by the first author of this paper (the only author that speaks the language that participants spoke in the video). As explained in the methodology section of this paper, the first author transcribed the videos. To avoid any possible mistakes, the first author watched every video at least twice. The second author of this paper then watched three randomly chosen videos, following the notes (written in English) to confirm that the notes reflected the observed behaviour.\footnote{Although the second author does not speak Brazilian Portuguese, many of the actions can be followed without the audio.} This validation step was meant to give us confidence that the video transcripts are a faithful description of the videos.

With the observation documents in hands, the first author then fully coded the document, which was then reviewed by the second author.
When designing this study, we discussed the possibility of a second researcher to code the transcribed version of the videos, as commonly done in qualitative research. We highlight the fact that the first author had a much more complete view of the data than the other two researchers. After all, although the transcripts of the videos that we created are factual and anyone can understand the actions of the participants by reading them, the videos are naturally a richer data source. While the first author was the one responsible for the initial coding, the derived framework was done through several discussions among the three researchers. The second and third author of the paper constantly challenged the analysis of the first author, which was then reflected back to the codes.

Even with all the checks, we agree that the analysis may reflect, even unconsciously, the first author's views on testing. For transparency, we make the readers aware that the first author is an expert in the area of software testing, and is aware of the current state-of-the-art tools, techniques, and practices being published in both the academic and in the technical fields. The first author has also authored two technical books on software testing, one about Test-Driven Development~\cite{aniche-book-1} and one about test automation~\cite{aniche-book-2}, and is currently responsible for the software testing course offered in the computer science bachelor at Delft University of Technology.\footnote{\url{https://studiegids.tudelft.nl/a101_displayCourse.do?course_id=55100}}

In addition, analyzing the participants' behaviour is a complex task as it involves understanding their concrete actions and thoughts that they verbally explain. We made sure to annotate only behaviours that were made explicit by either actions or words. This means that our results do not contain implicit actions or intentions that developers might have had in their minds, but that were never vocalized. 

Finally, while the 13 participants were recruited at once, at the end of the analysis, we believed we had achieved information saturation. We do not believe more participants would have provided us with more relevant data due to the very low number of new codes that were emerging while analyzing the last videos. In other words, we believe we have observed all different types of behaviors from the developers, at least the ones that could be triggered given our experimental setup. There was no need for more data collection. Nevertheless, as we recommended to researchers, replications of this work with a different set of participants and tasks may bring different perspectives on how developers engineer test cases.

The observation notes and the final codebook are available for inspection in our appendix~\cite{appendix}. Note that we do not share the raw videos of the participants as (i) we did not ask for their permissions, (ii) some videos contain private information (e.g., their e-mails or chat applications appear for a few seconds).

\subsection{External validity} External validity refers to the extent that the conclusions of the study apply to other contexts. In the case of this study, threats to the external validity are related to how much we can generalize the behaviour of our participants as well as the strategies and detailed explanations of their thought-processes to other software developers.

The 13 developers that participated in the observational study, as well as the 72 developers that took part in our survey have varied levels of experience in software development and in software testing. Moreover, the tasks we use in this study, while coming from a domain that is particularly known to developers, are complex enough to force developers to reason about testing. Nevertheless, these participants were selected by means of convenience sampling. Our original call for participation in the observational study tweet was retweeted by 80 different people and earned $\approx$24k impressions, and the call for participation for the survey was retweeted by 20 people and earned a total of $\approx$10k impressions.\footnote{According to Twitter Analytics, on February 27th, 2021.}
While these numbers indicate that the calls went much beyond our network bubble, we can not still rule out the possibility of participants being biased in ways we do not know. Moreover, all the tasks in the study are from a single domain (i.e., string manipulation), which may not be representative of all types of domains that developers write tests. Therefore, we do not argue our findings generalize to any developers working in any types of software systems and domains. Replications of this work are necessary before we can formulate a general theory on how developers test.

Finally, while we make several recommendations to developers, toolmakers and educators, we again highlight the fact that they are derived after what we observed in the 13 participants. Given the small sample size, one may argue that our recommendations may not be generalizable enough. We note that, although the sample size is small, we believe we have achieved saturation. In other words, we do not expect any behavior other than the ones we saw in our study, at least not with the same experimental tasks and goals. We also challenged our results through a survey with 72 practitioners. While we believe our recommendations are sound enough to be applied by developers, we remind readers that more replications are needed until we can be conclusive about them.

\section{Conclusion}

Engineering test cases is a challenge activity for software developers. In this paper, we aim at understanding the thought-process of developers when performing this activity. After observing 13 developers with varied levels of experience writing test cases for real-world open-source code and surveying 72 developers, we propose a framework and a set of strategies that explain how developers engineer test cases. Based on our observations, we suggest several actionable points for developers to improve the way they engineer test cases, tools that we believe would make developers more productive, and suggestions for educators on how to augment their testing courses. 

Our findings support developers in two ways: 
First, we provide a framework that can be used to formally explain how developers reason about engineering test cases. We hope our framework will support developers in better understanding how they engineer test cases and by understanding how others engineer test cases, and whether they see a way to improve their own practice.
Second, we show that some of the knowledge we already have as a field (especially in the academic community), such as the benefits of systematic testing and clear adequacy criteria, are not really applied by developers. We hope that our empirical findings serve as a first piece of evidence for developers to see the value of such practices.

Finally, while we believe this paper is the first of its kind when it comes to understanding the reasoning behind how developers devise test cases, it only paves the road towards a deeper understanding of the phenomenon. This paper suggests an extensive research agenda for empirical software engineering researchers that aim at improving the way developers test.

\section*{Acknowledgments}
We would like to thank the participants of our study. This work was partially sponsored by the Dutch science foundation NWO through the Vici “TestShift” project (No. VI.C.182.032). We gratefully acknowledge the support of the Swiss National Science Foundation through the SNF Projects No. 200021M\_205146

\bibliographystyle{IEEEtranN}
\bibliography{main}

\end{document}